\documentclass[aps,prb,english,twocolumn,showpacs,preprintnumbers,amsmath,amssymb,floatfix,superscriptaddress,longbibliography]{revtex4-1}
\usepackage[T1]{fontenc}
\usepackage[latin9]{inputenc}
\usepackage{graphicx}
\usepackage{amssymb}
\usepackage{babel}
\makeatletter
\usepackage[version=3]{mhchem}
\usepackage[usenames,dvipsnames]{xcolor}

\usepackage{ulem}

\global\arraycolsep=2pt
\usepackage{stmaryrd}
\usepackage{amsmath}
\usepackage{amssymb}
\usepackage{graphicx}
\usepackage{textcomp}
\usepackage{calrsfs}
\usepackage{yfonts}
\usepackage{bm}
\usepackage{color}
\newcommand{\ben}{\begin{equation*}}
\newcommand{\een}{\end{equation*}}
\newcommand{\bean}{\begin{eqnarray*}}
\newcommand{\eean}{\end{eqnarray*}}

\newcommand{\be}{\begin{equation}}
\newcommand{\ee}{\end{equation}}
\newcommand{\bea}{\begin{eqnarray}}
\newcommand{\eea}{\end{eqnarray}}

\newcommand{\psumbar}{\sum\nolimits^\prime}
\DeclareMathOperator*{\psum}{\psumbar}

\makeatother
\usepackage{babel}
\makeatother

\usepackage{color}

\usepackage{hyperref}
\hypersetup{
  colorlinks = true,
  citecolor  = blue,  
  linkcolor  = magenta 
}

\begin{document}
\title{A mechanism for ice growth on the surface of a spherical water droplet}

\author{Yang Li}
  \email{leon@ncu.edu.cn}
  \affiliation{School of Physics and Materials Science, Nanchang University, Nanchang 330031, China}
  \affiliation{Institute of Space Science and Technology, Nanchang University, Nanchang 330031, China}
\author{Prachi Parashar}
  \email{Prachi.Parashar@jalc.edu}
  \affiliation{John A. Logan College, Carterville, Illinois 62918, USA}
\author{Iver Brevik}
  \affiliation{Department of Energy and Process Engineering, Norwegian University of Science and Technology, NO-7491 Trondheim, Norway}
\author{Clas Persson}
  \affiliation{Department of Materials Science and Engineering, KTH Royal Institute of Technology, SE-100 44 Stockholm, Sweden}
\author{O. I. Malyi}
   \affiliation{Qingyuan Innovation Laboratory, Quanzhou 362801, China}
  \affiliation{Centre of Excellence ENSEMBLE3 Sp. z o. o., Wolczynska Str. 133, 01-919, Warsaw, Poland}
\author{Mathias Bostr{\"o}m}
  \email{mathias.bostrom@ensemble3.eu}
  \affiliation{Centre of Excellence ENSEMBLE3 Sp. z o. o., Wolczynska Str. 133, 01-919, Warsaw, Poland}
  \affiliation{Chemical and Biological Systems Simulation Lab, Centre of New Technologies, University of Warsaw, Banacha 2C, 02-097 Warsaw, Poland}

\begin{abstract}
The formation and growth of ice particles, particularly on the surfaces of spherical water droplets, bear profound implications for localized weather systems and global climate. Herein, we develop a theoretical framework for ice nucleation on minuscule water droplets, establishing that $10\sim5000\rm\ nm$ droplets can considerably increase in volume, making a substantial contribution to ice formation within mist, fog, or even cloud systems. We reveal that the Casimir-Lifshitz (van der Waals) interaction within these systems is robust enough to stimulate both water and ice growth on the surfaces of ice-cold spherical water droplets. The significant impacts and possible detectable phenomena from the curvature are demonstrated.
\end{abstract}

\date{\today}
\maketitle

\section{Introduction}

\par Small ice particles, an integral element of atmospheric science, for ice-phase processes in the global hydrological cycle and precipitation~\cite{MulenstadtSourdevalDalanoeQuaasGRL2015} are known to be important, which influences weather, as well as the global climate~\cite{StorelvmoTan2015}. The size and structure of ice nuclei can influence the radiation flux in the atmosphere~\cite{arienti2018experimental}, and thus weather and climate implications~\cite{zeng2009indirect,hawker2021temperature,burrows2022ice}. Ice nucleation is also of great importance in many other fields~\cite{maeda2021brief}.

\par Ice particles often measure less than a micrometer in diameter. These minute particles are born from a complex dance, where the water vapor adheres to and freezes on condensation nuclei diffusing throughout the cloud, fog, and mist. Consequently, it leads to the growth of intricate ice crystals. Our understanding on these phenomena is largely based on a rich tapestry of deposition theories, such as the well-established Wegener-Bergeron-Findeisen process~\cite{findeisen2015,StorelvmoTan2015}. At their core, these theories posit that the inherently unstable thermodynamic equilibrium between ice and water fosters the conversion of liquid clouds into ice clouds, facilitated by the preferential accretion of water vapor onto the ice at lower moisture concentrations. Within clouds, the ice nuclei~\cite{ValiIceNucleai_Nature1966}, known to initially be of the order $0.1\sim1\rm\,\mu m$, seem to require specific surface sites to induce heterogeneous ice growth~\cite{atkinson2013importance,Murraydoi:10.1126/science.aam5320,kiselev2017active}.
On occasions, observations of higher concentrations of ice particles than expected, have been addressed with various potential explanations~\cite{Lloydetalacp-20-3895-2020}: (i) presence of very efficient ice nucleating particles, (ii) recycling of ice in the downwelling mantle of the convective cloud,  and finally (iii) through a secondary ice production process.

\par In this contribution, we will present a model, where Casimir-Lifshitz energies (the van der Waals interactions), induced by the quantum vacuum and thermal fluctuations, can give rise to sufficiently large energies for hexagonal ice~\cite{AragonesMacDowellVega2011} to grow on the surface of ice-cold submicron-sized spherical water drops. Interestingly, it is energetically favorable for the continuous growth of the internal water droplet, and for each size of the inner liquid water region, an equilibrium is found, where a micron-sized ice layer forms on the outer surface. This study uses dielectric properties of water and ice corresponding to the triple point of water~\cite{JohannesWater2019,LUENGOMARQUEZMacDowell2021,LuengoMarquez_IzquierdoRuiz_MacDowell2022}, including its relevance to mist and fog formation at or near the ground level. However, it also means, in terms of cloud systems, the model is only valid for the given conditions and may not be applicable to all clouds.

\par It is Elbaum and Schick~\cite{Elbaum2,Elbaum} who firstly attempted to explain ice formation on a planar water surface by emphasizing the effect due to this Casimir-Lifshitz interaction, but for planar geometry. We focus, in the current work, on the new effects from this interaction, which has not previously been predicted, and can not be predicted in the quasi planar geometry. Also, it was not until improved water and ice dielectric functions became available that Fiedler et al.~\cite{JohannesWater2019}, and separately the team led by MacDowell~\cite{LuengoMarquez_IzquierdoRuiz_MacDowell2022}, predicted ice layer formation via dispersion interactions. Our work expands these works to consider the formation of ice layers on the surface of a water droplet. Notably, our prediction is that the inner surface of the ice layer (in contact with water) experiences partial premelting leading to the growth of the inner liquid water droplet, simultaneously as the outer ice surface (in contact with vapor) expands via the accumulation of water molecules from the surrounding moist. This is relevant, for instance, for water condensation in mist at conditions near the triple point of water. This is a prediction that comes out entirely from geometry considerations. It can not be understood from past works that focused on planar geometries. (At some point the inner surface of the ice layer solidifies and the growth will then only be on the outer ice surface where dispersion forces will promote further growth of ice layer until some equilibrium is reached).

\section{Models for ice and water}
\par The parameterized dielectric functions we utilize for water and ice~\cite{LUENGOMARQUEZMacDowell2021,LuengoMarquez_IzquierdoRuiz_MacDowell2022} agree very well (except for  at zero frequency) with the dielectric functions for discrete imaginary frequencies derived from experimental data on the real frequency axis, but they were evaluated in such a way that they do not obey the Kramers-Kronig relation (causality). In contrast, the water model by Fiedler, Parsons and co-workers~\cite{JohannesWater2019} was fitted directly on the real frequency axis and obeys causality. The zero-frequency dielectric constants for water and ice are adjusted based on experimental data. These data sets established on experiments~\cite{LUENGOMARQUEZMacDowell2021,LuengoMarquez_IzquierdoRuiz_MacDowell2022} present, unlike the data used in the past~\cite{Elbaum,Prachi_concentricice2019}, only one crossing between the curves for ice and water dielectric functions at imaginary frequencies. This will have far-reaching consequences for our prediction of a Wegener-Bergeron-Findeisen mechanism of ice growth in low-lying clouds, mists and fogs.
\begin{figure*}
  \centering
  \includegraphics[width=0.9\columnwidth]{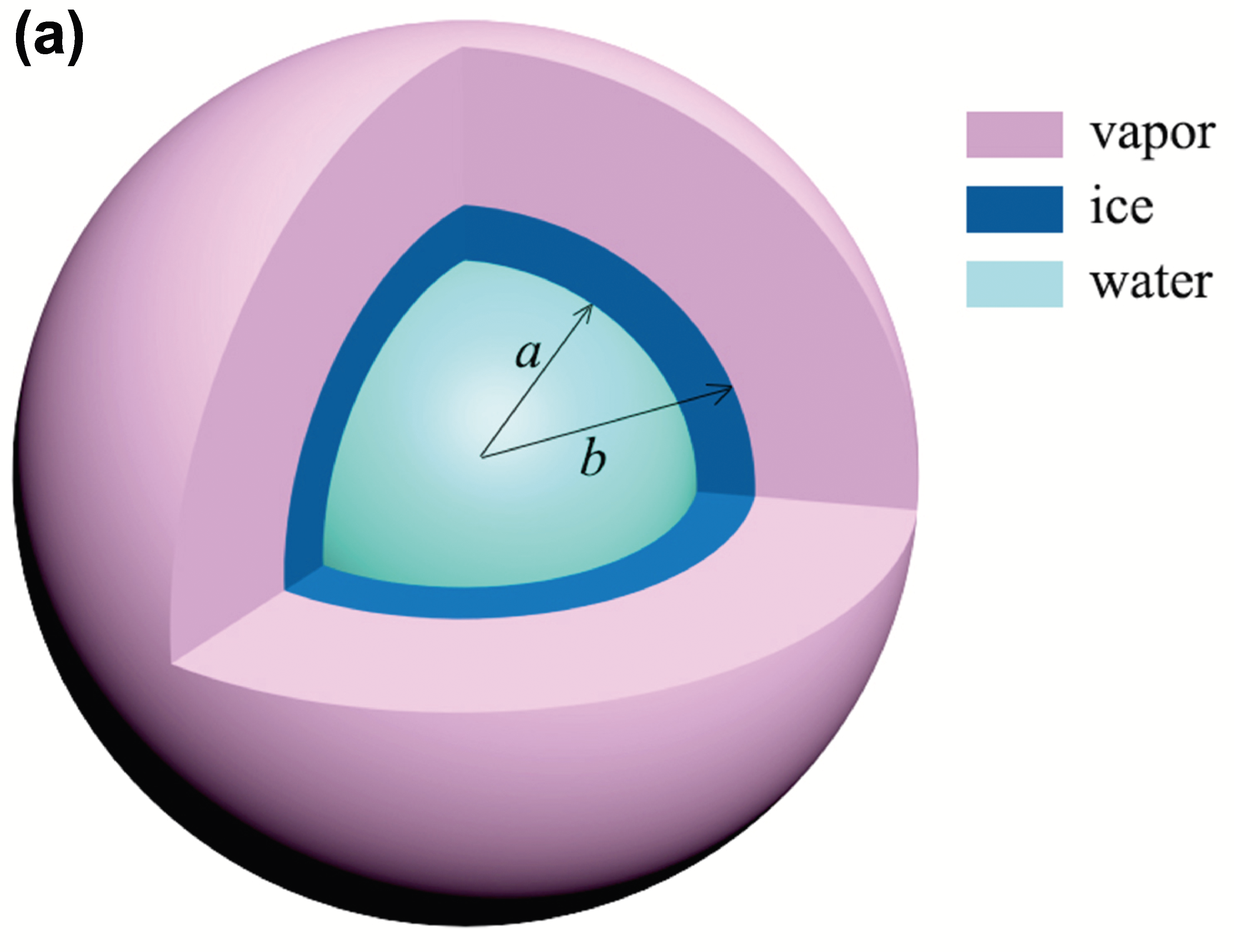}
  \includegraphics[width=1.1\columnwidth]{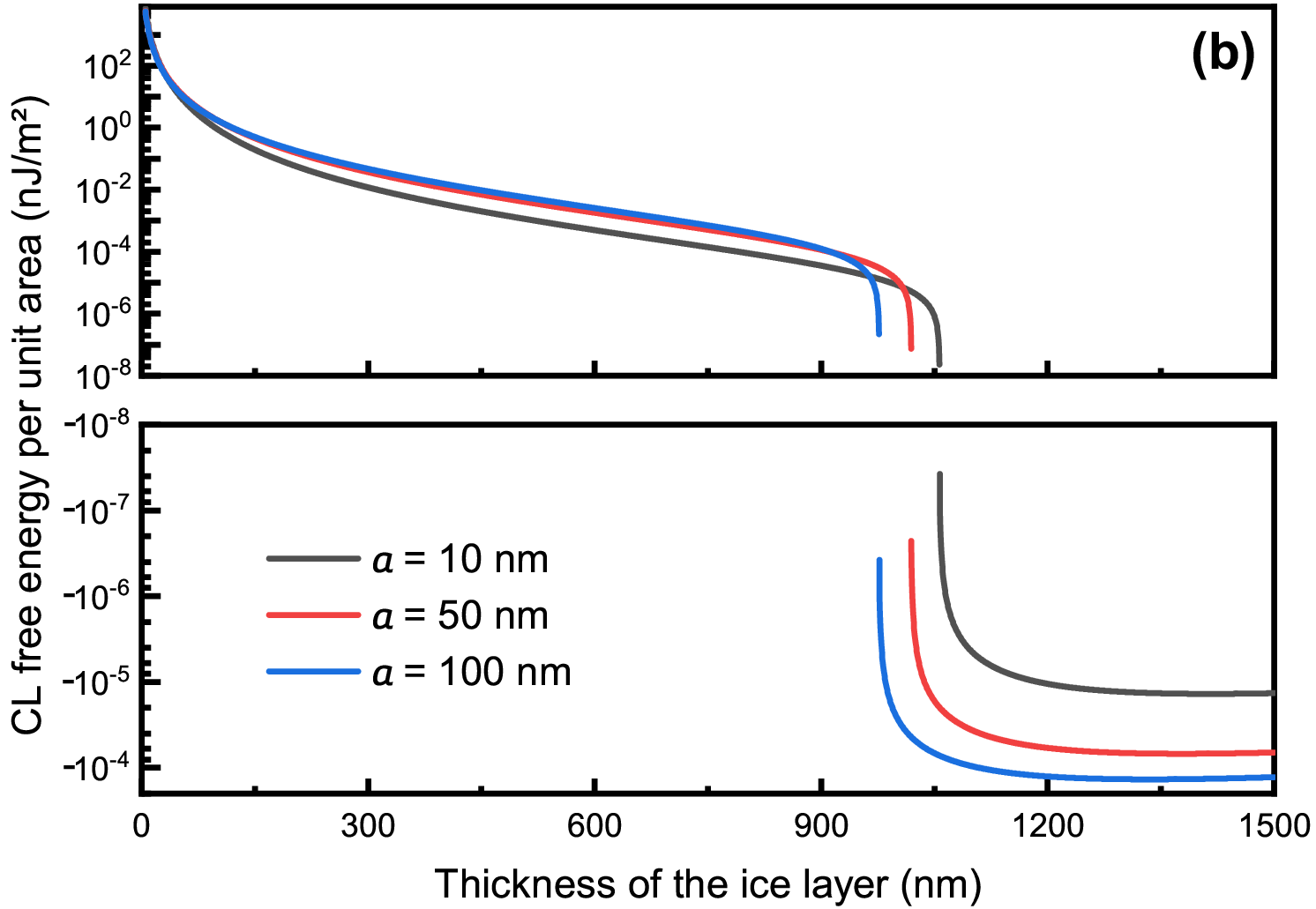}
  \caption{\label{fig:SchematicFigure}\textbf{(a)} The schematic illustration on the configuration studied, i.e., the water-ice-vapor spherical system, with the inner and outer radius of the ice layer being $a$ and $b$. \textbf{(b)} The Casimir-Lifshitz free energy per unit area as a function of the thickness of the ice layer, i.e., the difference between the outer $b$ and inner $a$ radius of the intervening layer in the water-ice-vapor concentric configuration.
  }
\end{figure*}

\section{Casimir-Lifshitz free energy for the concentric sphere geometry}
\par In Ref.~[\onlinecite{Prachi_concentricice2019}], Parashar, Shajesh, and their coworkers demonstrated that the variation of the Casimir-Lifshitz interaction (CL) energy, excluding the self-energy part, can be computed for layered concentric spherical ice particles. It should be mentioned that divergences usually plague the Casimir self-energy, including the bulk and surface contributions, of a single spherical body. This was made clear by K. A. Milton~\cite{milton1980semiclassical}, P. Candelas~\cite{candelas1982vacuum}, and analyzed by, for instance, I. Brevik et al.~\cite{brevik1983electromagnetic} some years ago. As shown by some of our authors, the divergences inside and outside the spherical surface are indeed cancelled out exactly, when the refractive indices on two sides of the surface are equal~\cite{brevik1983electromagnetic}. Similar argument is even applied to cylinder case~\cite{brevik1994casimir}. But typically this cancellation is not available. These divergences keep raising controversies up to recently~\cite{avni2018casimir,milton2020self}, and a satisfactory method to deal with these divergences is still in need~\cite{milton2018casimir,li2024casimir}. In this paper, however, we study the mechanism due to the Casimir-Lifshitz interaction between two media intermediated via a third medium, with two surfaces involved and no divergence occurring at all. The potential effects related to those divergences will be a novel topic for our studies in the near future.

\par For submicron-sized water droplets, we predict that the results are dramatically modified due to the curved spherical geometry. For full details of the theory for concentric sphere Casimir-Lifshitz free energy applied to ice-water systems, we refer to Ref.~[\onlinecite{Prachi_concentricice2019}] and the references therein. With the inner and outer ice layer radii labeled $a$ and $b$ as shown in Fig.~\ref{fig:SchematicFigure}a, the CL energy due to the transverse magnetic (TM) modes $F_{\rm H}$ is written as
\begin{eqnarray}
F_{\rm H}
&=&
T\sum_{n=0}^\infty{^\prime}\sum_{l=1}^\infty2\nu
\ln[1-r_{32}^{\rm H}(n,l;b)r_{21}^{\rm H}(n,l;a)],
\end{eqnarray}
where $\nu=l+1/2$, and the primed sum indicates that the first static term ($n=0$) should be weighted in half. The scattering coefficients are given by
\begin{subequations}
\label{eqTEco}
\begin{eqnarray}
r_{32}^{\rm H}(n,l;b)
&=&
\frac{
[e_l(\kappa_{3;n}b),e_l(\kappa_{2;n}b)]_{\varepsilon}
}{
[e_l(\kappa_{3;n}b),s_l(\kappa_{2;n}b)]_{\varepsilon}
},
\end{eqnarray}
\begin{eqnarray}
r_{21}^{\rm H}(n,l;a)
=
\frac{
[s_l(\kappa_{2;n}a),s_l(\kappa_{1;n}a)]_{\varepsilon}
}{
[e_l(\kappa_{2;n}a),s_l(\kappa_{1;n}a)]_{\varepsilon}
},
\end{eqnarray}
\end{subequations}
in which $\kappa_{i;n}=2\pi nT\sqrt{\varepsilon_i\mu_i},\ i=1,2,3$ (with the indices corresponding to water, ice, vapor respectively), the temperature $T=273.16\rm\ K$, and the bracket $[\cdot,\cdot]_{\varepsilon}$, involving the permittivity $\varepsilon$, is defined as
\begin{equation}
[f(x),g(x)]_{\varepsilon}=\frac{f'(x)}{\varepsilon_f}g(x)-f(x)\frac{g'(x)}{\varepsilon_g}.
\end{equation}
$e_l(x)=\sqrt{2x/\pi}K_{\nu}(x)$ and $s_l(x)=\sqrt{\pi x/2}I_{\nu}(x)$, with $K_{\nu}(x)$ and $I_{\nu}(x)$ being the modified spherical Bessel functions. The transverse electric (TE) CL free energy, which will be studied as the representative quantity in the Appendix,
is obtained with the substitutions $H\rightarrow E$, $\mu\leftrightarrow\varepsilon$. For nonmagnetic dielectrics, the $n=0$ terms can be explicitly written as $F_{\rm E}^{n=0}=0$ and
\begin{equation}
F_{\rm H}^{n=0}=T\sum_{l=1}^{\infty}\nu
\ln\bigg[
1+\frac{
(\varepsilon_3-\varepsilon_2)(\varepsilon_2-\varepsilon_1)
}{
(\frac{l+1}{l}\varepsilon_3+\varepsilon_2)
(\varepsilon_2+\frac{l}{l+1}\varepsilon_1)
}
\bigg(\frac{a}{b}\bigg)^{2\nu}
\bigg],
\end{equation}
where $\varepsilon_i$ takes its static value. The impacts of the spherical geometry are evident even for the relatively simple static part, implying their significance as shown in the following limiting cases.

\par Suppose the separation between inner and outer interfaces $d=b-a$ goes to zero with $a$ remaining constant. Then in this limit, the retardation effects are too trivial to be included, and only non-retarded contributions survive. The TM CL energy, as an instance, in this case is
\begin{eqnarray}
F_{\rm H}
=
-4\pi a^2\frac{A}{12\pi d^2}
,\ d\rightarrow0,
\end{eqnarray}
where $A$ is the Hamaker constant of the corresponding planar three-layer configuration. (See the Appendix for detailed calculation of TM CL energy.) We thus claim that in this small-separation limit, the CL interaction in the concentric configuration is just the same as that in the planar case. It is not a surprising result, since at a very small separation the local interaction dominates, meaning that the impacts due to the geometry are negligible. However, it happens only when the separation is extremely small. Generally within the scale where the continuum theory is valid, the retardation effects are highly nontrivial, as shown in the Appendix. Therefore, the retardation, even in our case where the size of the configuration may look small, should play an important role.

\par Even if the retardation is fully counted, the geometry of the configuration acts as another factor making the situation more complicated. In contrast with the small-separation limit above, if $d=b-a$ kept constant but $a\rightarrow\infty$ (referred to as the ``planar limit'' here), then TM Casimir-Lifshitz free energy $F_{\rm H}$, based on the Appendix, behaves as
\begin{eqnarray}
\label{eq.FEalarge}
\frac{F_{\rm H}}{4\pi a^2}
&=&
T\psum_{n=0}^{\infty}\int_{0}^{\infty}\frac{dkk}{2\pi}\ln\bigg(
1
+
\frac{
\varepsilon_2\widetilde{\kappa}_{3,n}-\varepsilon_3\widetilde{\kappa}_{2,n}
}{
\varepsilon_2\widetilde{\kappa}_{3,n}+\varepsilon_3\widetilde{\kappa}_{2,n}
}
\nonumber\\
& &
\times
\frac{
\varepsilon_1\widetilde{\kappa}_{2,n}-\varepsilon_2\widetilde{\kappa}_{1,n}
}{
\varepsilon_1\widetilde{\kappa}_{2,n}+\varepsilon_2\widetilde{\kappa}_{1,n}
}e^{-2\widetilde{\kappa}_{2,n}d}
\bigg),
\end{eqnarray}
in which $\widetilde{\kappa}_{i,n}=\sqrt{k^2+\varepsilon_i\zeta_n^2}$, and the famous Dzyaloshinskii-Lifshitz-Pitaevskii (DLP) formula for the CL free energy per unit area of the planar three-layer configuration~\cite{lifshitz1956theory,dzyaloshinskii1961general} is exactly derived. The same arguments apply to the TE contribution. According to Eq.~\eqref{eq.FEalarge}, the impacts due to the geometry, which are embodied in the curvature characterized by the factor $a$, vanish as $a$ approaches infinity. This result is consistent with the assumption usually employed, that is, for a relatively large medium, the Casimir-Lifshitz interaction near its surface can be evaluated in terms of the planar configuration. As directly shown below, the effects of curvature here can only be fully eliminated by decreasing the curvature, provided the separation between interfaces are not extremely small. In this sense, our analysis demonstrates the importance of understanding curvature effects for submicron-layered systems.

\par Given that the analytical investigation as above is not generally available because of the complex forms of permittivities of real materials, we mainly resort to numerical evaluations to unveil corresponding phenomena and physics. Results will be presented for the concentric sphere geometry with inner radii ranging from $10\rm\ nm$ to $1\ \rm\mu m$, exactly in the size range of observed ice nuclei discussed in the introduction.

\section{Results and Discussions}

\par It was recently predicted that minimization of Casimir-Lifshitz free energies could induce ice layer formation at the surfaces of water~\cite{JohannesWater2019,LUENGOMARQUEZMacDowell2021,LuengoMarquez_IzquierdoRuiz_MacDowell2022}, as well as on AgI particles in clouds~\cite{LUENGOMARQUEZMacDowell2021}. The mechanism for ice formation has also been invoked for self-preservation of gas hydrates~\cite{bostrom2021self}.
However, all those works, predicting the minimization of free energy with the ice growth, exploited theories with the planar approximation, which in the submicron size, are often better described as concentric spherical systems. The study on energies in the actual submicron concentric spherical systems unveils interesting new physics. The CL free energy curves as functions of the ice layer thickness, for water droplets with given inner radii ($a=10,\ 50$, and $100\rm\ nm$), are shown in Fig.~\ref{fig:SchematicFigure}b. Two things are of immediate interest. Firstly, the depth of the minimum for the free energy per unit area increases, as the water droplet radius increases.  It follows from energy considerations that close to micron-thick ice layers grow on nano-sized water droplets under the right conditions. Thereafter, the inner liquid core of the layered droplet grows at the expense of the inner parts of the ice coating. The ice film thickness, in turn, is kept close to constant by water vapor adsorption, leading to an expanding inner radius ($a$) and thus the radius of the composite nuclei ($b$). Notably, with a thinner ice layer, it is not energetically favorable for the water droplet to grow since the energy increase with increasing $a$ for small ice layer thicknesses. The second important observation is that the relative amount of ice in the droplet, which minimizes the energy, increases with decreasing water droplet radius. As can be seen in Table~\ref{table1}, the relative amount of ice growth is predicted to be dramatic for nano-sized water droplets. This leads us ultimately to predict an enhanced ice growth mechanism for ice nucleating nano-droplets. This occurs in a new ``secondary'' way, via a reduced CL free energy, that leads to the simultaneous growth of both the interior water droplet and the ice coating with the adsorption of water molecules from the vapor phase. This results in larger volumes of ice, but should not be mixed up with the traditional secondary ice processes. Furthermore, the planar approximation, compared with a water droplet of a radius 10 nm, gives a correct order of magnitude estimate for the predicted ice coating thickness. It should be emphasized that for a fixed water droplet size, the repulsive Casimir-Lifshitz stress can be substantially increased below the equilibrium ice film thickness. This can lead to the establishment of the initial growth of an ice layer due to dispersion energies.

\par
\begin{figure*}
  \centering
  \includegraphics[scale=0.35]{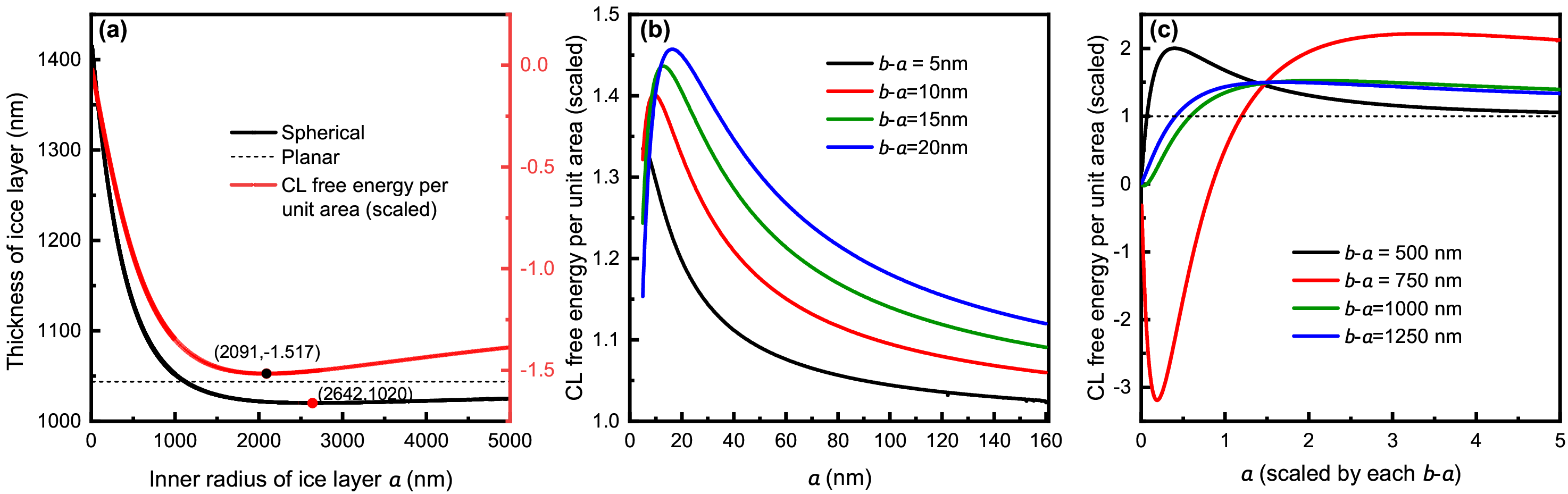}
  \caption{\label{fig:CasLifFreeEnergyNormalized} \textbf{(a)} The thickness of the ice layer (solid line) minimizing the CL free energy per unit area as a function of droplet's inner radius $a$. The global minimum thickness is about $1020\rm\ nm$ located near $a=2642\rm\ nm$ (red dot). {The minimum thickness of the ice layer in the corresponding planar configuration is also shown (dotted line) as a reference. The minimum CL free energy per unit area for each $a$, scaled by the absolute value of that for the planar case, are plotted as the red line, with the black dot being the minimum.} \textbf{(b)} With small ice layer thicknesses, the CL free energy per unit area of the concentric configuration, scaled by their corresponding planar cases with the same ice layer thicknesses, as the function of $a$. \textbf{(c)} The plot is in the same manner as \textbf{(b)} except for the relatively large ice layer thicknesses.}
\end{figure*}

 Moreover, we see a minimum equilibrium thickness of ice layer, when varying the inner radius $a$, as demonstrated by Fig.~\ref{fig:CasLifFreeEnergyNormalized}a. The red dot there marks the minimum thickness of ice layer at $1020\rm\ nm$ with $a$ being about $2642\rm\ nm$. However, the minimum of the Casimir-Lifshitz free energy per unit area, shown as the black dot in Fig.~\ref{fig:CasLifFreeEnergyNormalized}a, deviates from the minimum thickness. This small deviation is due to the combination of the size of the sphere and the free energy per unit area.  It should be noted that the CL free energy of the spherical system, instead of the CL free energy per unit area as in planar cases, should be minimized to achieve the stability of the system. The total Casimir-Lifshitz free energy in this system does not show a minimum. The CL free energy per unit area here, which is defined as the total free energy devided by the area of the inner interface, is employed to compare with the corresponding planar case, while the CL free energy itself decreases monotonically as the size of the sphere increases, since the planar case is approached then. (See Table~\ref{table1} for numbers showing this tendency.)
\begin{table}
\centering
\begin{tabular}{c|c|c|cc}
  \hline
   $a$ (nm) & ice thickness (nm) & $F_{\rm min}$ (nJ/m$^2$) & volume ratio ($\%$) \\
   \hline
   10 & 1415 & $-1.38\times10^{-5}$ & $2.89\times10^8$ \\ 
   \hline
   20 & 1405 & $-2.75\times10^{-5}$ & $3.62\times10^7$ \\ 
   \hline
   40 & 1386 & $-5.50\times10^{-5}$ & $4.53\times10^6$ \\ 
   \hline
   60 & 1368 & $-8.23\times10^{-5}$ & $1.35\times10^6$ \\ 
  \hline
   80 & 1351 & $-1.09\times10^{-4}$ & $5.72\times10^5$ \\ 
    \hline
   100 & 1334 & $-1.36\times10^{-4}$ & $2.94\times10^5$ \\  
    \hline
   200 & 1262 & $-2.66\times10^{-4}$ & $3.90\times10^4$ \\  
          \hline
   400 & 1163 & $-4.88\times10^{-4}$ & $5.87\times10^3$ \\  
             \hline
   600 & 1106 & $-6.51\times10^{-4}$ & $2.20\times10^3$ \\  
         \hline
   800 & 1072 & $-7.61\times10^{-4}$ & $1.18\times10^3$ \\  
        \hline
   1700 & 1025.18 & $-9.31\times10^{-4}$ & $3.12\times10^2$ \\  
        \hline
   2500 & 1020.18 & $-9.34\times10^{-4}$ & $1.79\times10^2$ \\  
        \hline
   3000 & 1020.39 & $-9.20\times10^{-4}$ & $1.41\times10^2$ \\  
        \hline
   3500 & 1021.32 & $-9.04\times10^{-4}$ & $1.16\times10^2$ \\  
        \hline
   4000 & 1022.52 & $-8.88\times10^{-4}$ & $9.80\times10^1$ \\  
        \hline
   5000 & 1025.26 & $-8.58\times10^{-4}$ & $7.50\times10^1$ \\  
        \hline
   5500 & 1026.14 & $-8.45\times10^{-4}$ & $6.71\times10^1$ \\  
        \hline
  $\infty$ & 1043.84 & $-6.19\times10^{-4}$ & ---\\
  \hline
\end{tabular}
\caption{\label{table1} The equilibrium ice layer thickness at minima for the free energy per unit area $F_{\rm min}$ (for inner surface of ice layer) as a function of water droplet radius ($a$).
We also present in the table the droplets' relative volume growth due to ice layer formation in percentage ($100\times(b^3-a^3)/a^3\ \%$).
}
\end{table}

\par Although in the considered range of water droplet sizes, the CL free energy is minimized for ice layer thicknesses of same order of magnitude as was predicted for a planar water surface~\cite{JohannesWater2019,LUENGOMARQUEZMacDowell2021,LuengoMarquez_IzquierdoRuiz_MacDowell2022}, the impact of curvature demonstrates itself by the variation of the ice layer thickness within several hundreds of nanometers. In Fig.~\ref{fig:CasLifFreeEnergyNormalized}b, we demonstrate that for relatively thin ice layers, the CL free energy per unit area enhanced due to the curvature approaches the planar approximation for $a\geq100\rm\ nm$. The enhancement is weaker for thinner ice layer. Besides, with the varying $a$, the enhancement with respect to its planar counterpart is not monotonic, since the interaction between the inner and outer surfaces of the ice layer drastically decays as the size of water ball goes to zero. For thicker ice layers, as shown in Fig.~\ref{fig:CasLifFreeEnergyNormalized}c, more complicated dependencies on $a$ are observed. When the ice layer is still relatively thin, $b-a=0.5\rm\ \mu m$ , similar behavior as in Fig.~\ref{fig:CasLifFreeEnergyNormalized}b is seen. With larger thicknesses, the speed of converging to the planar case is typically small. In addition, a dip, which is particularly apparent for $b-a=0.75\rm\ \mu m$, appears, since the CL free energy of planar case for scaling changes sign. The curvature tends to raise the CL free energy, but if the surfaces are too far away from each other and the inner surface is too small, a strong enough interaction can hardly be maintained.


\par Table~\ref{table1} also shows us the predicted decreasing ice-water ratio, as the radius of the inner surface of ice layer increases, which implies a particularly noticeable growth-promoting effect for small water drops. This result, in the larger perspective, displays a ``secondary'' ice growth mechanism that should be important enough to have meteorological impacts. As demonstrated in the first half of the 20th century~\cite{findeisen2015,StorelvmoTan2015} by Wegener, Bergeron, and Findeisen, mixtures of ice and water are thermodynamically unstable.  Here, our prediction is that the interior water droplet is in contact with a micron-sized ice layer.  Therefore, in subsequent steps, the inner, unfrozen, water droplet core (in contact with ice) will eventually freeze, but the droplet ends up substantially larger compared to the case when only the initial water droplet had frozen.

\par Notably, Mie scattering of light against ice-coated water droplets depends on both the relative amounts of water and ice, as well as or more importantly, the nuclei sizes. This means the predicted effect is already part of our daily life, when we ponder the colors of the sky. A very useful way of measuring particle size is to make use of Mie scattering. Consider a plane wave of intensity $I_0$ falling towards a uniform dielectric sphere of radius $a$ and refractive index $n$. The Mie scattering formalism is valid for all values of $b$ relative to the wavelength $\lambda$ of the incident light, but we will here confine ourselves to the Rayleigh approximation, which holds when the size of the sphere is much smaller than the wavelength. In practice, this means that the Rayleigh approximation is useful for $b<\lambda/10$, or approximately $b<50\rm\ nm$ in optics. Infrared radiation, notably, is between 780\ nm and 1 mm. Then, the intensity $I$ observed at a distance $R$ away from the scatterer is
\begin{equation}
I= I_0\frac{1+\cos^2\theta}{2R^2}\left( \frac{2\pi}{\lambda}\right)^4
\label{intensity}
\end{equation}
where $\theta$ is the scattering angle. This expression shows that $I$ is very sensitive versus variations in size of the layered water droplet. As the system grows, this leads to many orders of magnitude corrections for the scattered intensity. A more accurate approach would be to use Mie (or Rayleigh) theory for stratified fluid spheres~\cite{Kerker}, which makes it possible to incorporate the difference between the refractive indices for ice and water. In practice, this additional refinement would hardly be of any importance, however, as these refractive indices are so similar.

\section{Conclusions}

\par Herein, we yield novel insights into coupled ice formation and premelting processes, emphasize the crucial role of accurate material parameterization and geometry, and shed light on the role of Casimir-Lifshitz interactions. As shown in the current letter, within appropriate geometry and using energy considerations, micron-thick ice layers grow on nano-sized spherical water droplets under the right conditions. Thereafter we predict the inner liquid core of the layered droplet to grow at the expense of the inner parts of the ice coating. The ice film thickness, in turn, is kept close to constant by the condensation, and subsequent freezing, of water vapor at the outer ice surface. The thermodynamic instability inherent in ice-water mixtures then leads to the eventual freezing of the interior unfrozen water droplet. The resulting ice droplet is substantially larger than predicted if solely the initial water droplet had frozen, challenging conventional understandings of ice formation. A noteworthy aspect of this mechanism is the Mie scattering of light against ice-coated water droplets, which is directly impacted by the relative proportions of water/ice and the size of the nuclei. Intriguingly, this effect permeates into everyday life, subtly influencing the colors we see. These findings unsettle previous predictions and authenticate the formation of a micron-sized ice layer on the surface of cold water, a stark contradiction to prior assumptions of premelting. Recently, a complex spectrum of interactions~\cite{Esteso4layerPCCP2020} and stability states in our exploration of planar three-layer and four-layer scenarios~\cite{LiMiltonBrevikMalyiThiyamPerssonParsonsBostrom_PRB2022} are discovered. Together with the important influences from the geometry, diverse ice premelting and formation phenomena in more complicated systems are anticipated.

Our analysis does operate on certain assumptions, such as the insignificance of surface charge and surface adsorption of free ions~\cite{ThiyamFiedlerBuhmannPerssonBrevikBostromParsons2018}. At low salt concentrations and pH around 7, this is expected to be a useful approximation, but may gloss over some potentially pivotal aspects of ice formation and premelting processes (e.g. pH effects from sulfur in the atmosphere from volcanoes~\cite{STOIBER1987}), indicating the need for further research to decipher charge-induced interactions~\cite{ParsonsSalis2015,Parsons2022}. We also point out that the interfacial tension of the droplet is related to attractive forces in the interface region, which are only partially encompassed by the van der Waals energies. There would be effects from the size-dependence of surface tension of the curved water-vapor interface. Pruppacher and Klett gave a quasi-thermodynamical relation  between radius and changes in surface tension~\cite{PruppacherKlett},
\begin{equation}
\sigma_{w/v}={\frac{\sigma_{w/v}(\infty)}{1+(2/a)\Gamma_w^{w/v}/(\rho_w-\rho_v)}},
    \label{Pruppacher}
\end{equation}
where $\sigma_{w/v}$ is water-vapor surface tension at radius $a$, $\sigma_{w/v}(\infty)$ the surface tension for a planar water-vapor interface, $\Gamma_w^{w/v}$ the Gibbs adsorption, and $\rho_w-\rho_v$ is the difference in the mass densities of water and vapor. Future research should also address various other related factors, including the dynamics of water molecules, to constrain the relevant nucleation parameters~\cite{KnopfAlpertNatureRevPhys2023}.


\section*{Appendix: Limiting behaviors}
\label{App.ICC}
\par Suppose the separation between the inner and outer interface $d=b-a$ goes to zero, while $a$ is kept constant. Obviously, in the zero-separation case, $d=0$, the Casimir free energy in the concentric configuration diverges, as the corresponding case in the planar system. To capture its small-separation behaviors, we consider the separation nonzero, but infinitely small, that is, the TM Casimir free energy per unit area with respect to the inner interface, for instance (TE part following the same arguments), expressed as
\begin{subequations}
\begin{eqnarray}
f_{\rm H,s}
&=&
\frac{T}{4\pi a^2}\sum_{n=0}^\infty{^\prime}\sum_{l=1}^\infty2\nu
\ln[1-r_{32}^{\rm H}(n,l;b)r_{21}^{\rm H}(n,l;a)],\quad\quad
\end{eqnarray}
in which $b=a+d$, $\nu=l+1/2$, functions $e_l(x)$ and $s_l(x)$ are defined and utilized as in the main text, and
\begin{eqnarray}
r_{32}^{\rm H}(n,l;b)
&=&
\frac{
[e_l(\kappa_{3;n}b),e_l(\kappa_{2;n}b)]_{\varepsilon}
}{
[e_l(\kappa_{3;n}b),s_l(\kappa_{2;n}b)]_{\varepsilon}
},
\end{eqnarray}
\begin{eqnarray}
r_{21}^{\rm H}(n,l;a)
=
\frac{
[s_l(\kappa_{2;n}a),s_l(\kappa_{1;n}a)]_{\varepsilon}
}{
[e_l(\kappa_{2;n}a),s_l(\kappa_{1;n}a)]_{\varepsilon}
},
\end{eqnarray}
\end{subequations}

\par In the small-separation limit $d\rightarrow0_+$, the $n=0$ contributions are expressed as
\begin{equation}
\label{eq.FH0d0}
\widetilde{f}_{\rm H,s}^{n=0}\sim \frac{T}{4\pi a^2}\sum_{l=1}^{\infty}\nu\ln\bigg[
1+\frac{
(\varepsilon_3-\varepsilon_2)(\varepsilon_2-\varepsilon_1)e^{-\frac{2\nu}{a}d}
}{
(\frac{l+1}{l}\varepsilon_3+\varepsilon_2)
(\varepsilon_2+\frac{l}{l+1}\varepsilon_1)
}
\bigg],
\end{equation}
where the leading order in $(a/b)^{2\nu}=e^{-2\nu\ln(1+d/a)}\approx e^{-2\nu d/a}$ for arbitrarily $\nu$ is used, since $d/a\rightarrow0$ is the smallest dimensionless scale involved here. For the planar limit, the Casimir free energy per unit area due to the $n=0$ contribution is written as
\begin{eqnarray}
\widetilde{f}_{\rm H,p}^{n=0}
&\sim&
\frac{T}{2}\int_{0}^{\infty}\frac{dkk}{2\pi}\ln\bigg[
1+\frac{
(\varepsilon_3-\varepsilon_2)(\varepsilon_2-\varepsilon_1)
}{
(\varepsilon_3+\varepsilon_2)
(\varepsilon_2+\varepsilon_1)
}
e^{-2kd}
\bigg]
\nonumber\\
&=&
-\frac{A_0}{12\pi d^2}
,
\end{eqnarray}
where $A_0$ is the $n=0$ Hamaker constant of the corresponding planar three-layer configuration. It is demonstrated that for a general inner radius $a$, $\widetilde{f}_{\rm H,s}^{n=0}$ behaves differently from its planar counterpart. The most evident deviations are seen for relatively small angular index $l$, and as $l$ increases, the corresponding planar results are approached, since the differences between $\frac{l+1}{l}\varepsilon_3,\ \frac{l}{l+1}\varepsilon_1$ and $\varepsilon_3,\ \varepsilon_1$ in the denominator in Eq.~\eqref{eq.FH0d0} vanish.

\par For $n>0$ contributions, the similar behaviors are observed. For planar case, in the small-separation limit, where the retardation is not significant, the contribution from the $n$th ($n>0$) order of Matsubara frequency to the Casimir free energy per unit area is expressed as
\begin{eqnarray}
\label{eq.fnHp}
\widetilde{f}^n_{\rm H,p}
&=&
\frac{T}{2\pi d^2}\int_0^{\infty}dkk\ln
\bigg[
1+R_{321}^{(p)}(n)e^{-2k}
\bigg]
.
\end{eqnarray}
where the factor $R_{321}^{(p)}(n)$ is
\begin{eqnarray}
R_{321}^{(p)}(n)=
\frac{\varepsilon_3(i\zeta_n)-\varepsilon_2(i\zeta_n)}{\varepsilon_3(i\zeta_n)+\varepsilon_2(i\zeta_n)}
\frac{\varepsilon_2(i\zeta_n)-\varepsilon_1(i\zeta_n)}{\varepsilon_2(i\zeta_n)+\varepsilon_1(i\zeta_n)}
.
\end{eqnarray}
In terms of the similar form, we can write the contribution for the concentric configuration corresponding to Eq.~\eqref{eq.fnHp}, in the small-separation limit, as
\begin{eqnarray}
\label{eq.fnHs}
\widetilde{f}^n_{\rm H,s}
&=&
\frac{T}{2\pi d^2}\sum_{l=1}^{\infty}\frac{\nu d^2}{a^2}\ln\bigg[
1+R_{321}^{(s)}(n,l;a)e^{-\frac{2\nu}{a}d}
\bigg]
,
\end{eqnarray}
where the correspondence $d/a\rightarrow dk,\ \nu d/a\rightarrow k$ as above can be applied. To explicitly figure out $R_{321}^{(s)}(n,l;a)$, it is observed that large orders of $l$ contribute most significantly, which means the uniform asymptotic expansion (UAE) can be employed~\cite{olver2010nist}. Then the functions $e_l(x)$ and $s_l(x)$ are approximated as
\begin{subequations}
\begin{eqnarray}
e_l(x)\sim\frac{\sqrt{z}e^{-\nu\eta(z)}}{(1+z^2)^{\frac{1}{4}}},\ s_l(x)\sim\frac{\sqrt{z}e^{\nu\eta(z)}}{2(1+z^2)^{\frac{1}{4}}},
\end{eqnarray}
\begin{eqnarray}
e'_l(x)\sim\frac{e^{-\nu\eta(z)}(1+z^2)^{\frac{1}{4}}}{-\sqrt{z}},\ s'_l(x)\sim\frac{e^{\nu\eta(z)}(1+z^2)^{\frac{1}{4}}}{2\sqrt{z}},\quad\quad
\end{eqnarray}
where $z$ and $\eta(z)$ are $z=x/\nu$ and
\begin{eqnarray}
\eta(z)=\sqrt{1+z^2}+\ln\frac{z}{1+\sqrt{1+z^2}}.
\end{eqnarray}
\end{subequations}
We are thus led to ($z_{i,x}=\sqrt{\varepsilon_i}\zeta_{n}x/\nu$)
\begin{subequations}
\begin{equation}
r_{32}^{\rm H}(n,l;b)
\sim
\frac{
\varepsilon_2(1+z_{3,b}^2)^{1/2}
-
\varepsilon_3(1+z_{2,b}^2)^{1/2}
}{
\varepsilon_2(1+z_{3,b}^2)^{1/2}
+
\varepsilon_3(1+z_{2,b}^2)^{1/2}
}\frac{2}{e^{2\nu\eta(z_{2,b})}},
\end{equation}
\begin{equation}
r_{21}^{\rm H}(n,l;a)
\sim
\frac{
\varepsilon_1(1+z_{2,a}^2)^{1/2}
-
\varepsilon_2(1+z_{1,a}^2)^{1/2}
}{
\varepsilon_1(1+z_{2,a}^2)^{1/2}
+
\varepsilon_1(1+z_{1,a}^2)^{1/2}
}\frac{e^{2\nu\eta(z_{2,a})}}{-2},
\end{equation}
\end{subequations}
which result in the $-r_{32}^{\rm H}(n,l;b)r_{21}^{\rm H}(n,l;a)$ with $d\rightarrow0_+$ expressed as
\begin{eqnarray}
-r_{32}^{\rm H}(n,l;b)r_{21}^{\rm H}(n,l;a)
\sim
\frac{
\varepsilon_2-\varepsilon_3
}{
\varepsilon_2+\varepsilon_3
}
\frac{
\varepsilon_1-\varepsilon_2
}{
\varepsilon_1+\varepsilon_1
}e^{-2\nu d/a}
.
\end{eqnarray}
Therefore, in the small-separation limit, $R_{321}^{(s)}(n,l;a)$ approaches $R_{321}^{(p)}(n)$, and the distinction between the spherical and planar geometries disappears. Evidently, the same arguments applies to the TE contributions.

\begin{figure}
  \centering
  \includegraphics[width=1.0\columnwidth]{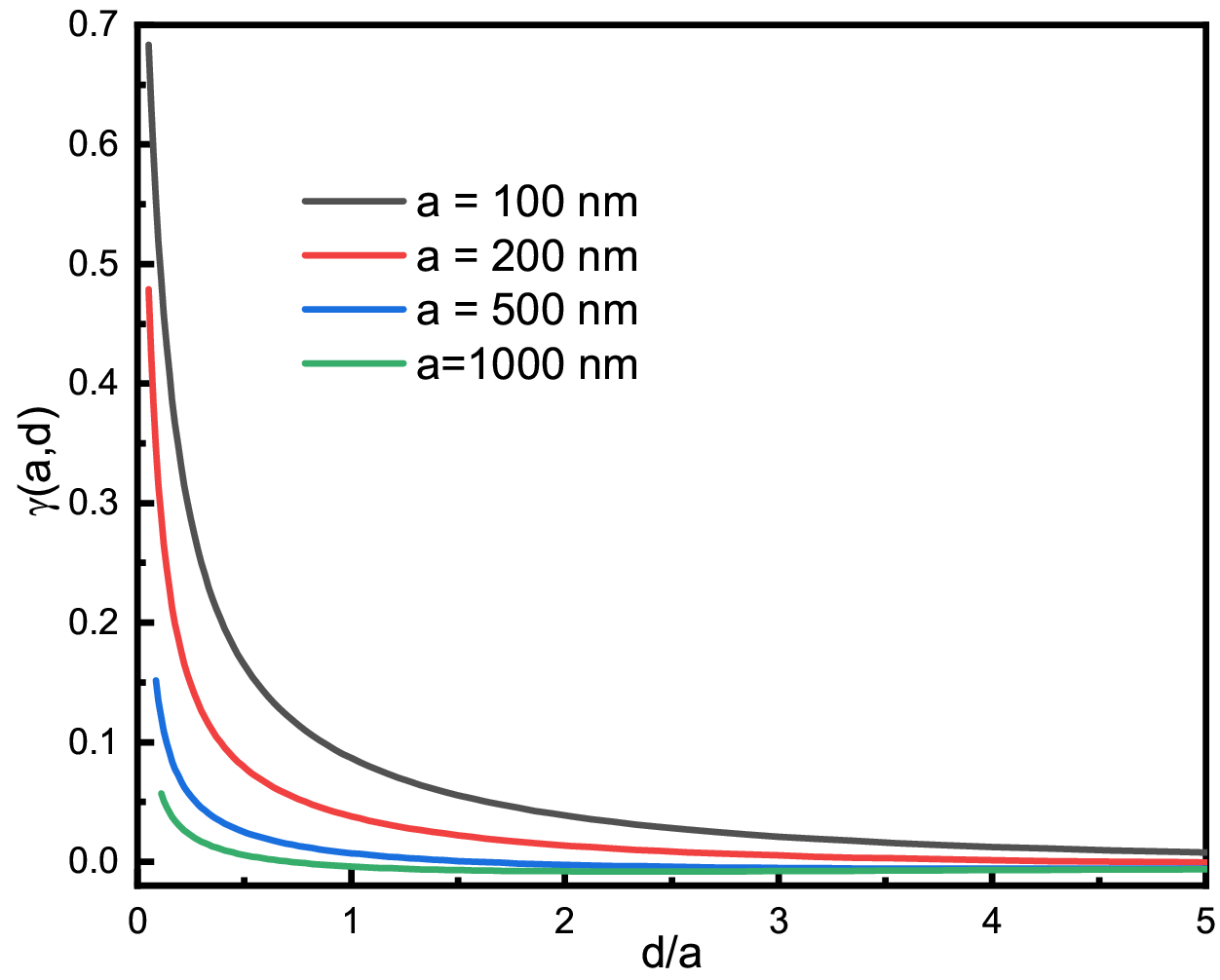}
  \caption{\label{fig:Hamaker}The factor $\gamma(a,d)$ defined in Eq.~\eqref{eq.Hamaker1} as a function of $d/a$ for each given inner radius $a=100{\rm nm}$ (black), $a=200{\rm nm}$ (red), $a=500{\rm nm}$ (blue), and $a=1000{\rm nm}$ (green).}
\end{figure}
\par As a glimpse of the importance of the retardation, we compare the Casimir free energy per unit area for the water-ice-vapor concentric configuration, with the corresponding small-separation value where the retardation can be ignored, that is,
\begin{eqnarray}
\label{eq.Hamaker1}
\gamma(a,d)\equiv
-\frac{12\pi d^2 f_{\rm H,s}(a,d)}{A},
\end{eqnarray}
in which $A$ is the total Hamaker constant. In Fig.~\ref{fig:Hamaker}, $\gamma(a,d)$ is shown for some fix inner radius. When $d/a$ is large, meaning that the separation $d$ can not be considered small in any sense, the discrepancy between the retarded and non-retarded free energy is manifest. Even $d/a$ approaches to a relatively small value about $0.05$, the distinction is still obvious, especially when $a$ is large. This result is natural, since the retardation can be ignored only if the absolute value of the separation is small. Therefore, even the concentric system might be tiny (in the hundreds or even tens of nanometer scale), the retardation can hardly be ignored.

\par On the other way round, when $d=b-a$ kept constant but $a\rightarrow\infty$, we arrive at another limitting case (or referred to as the ``planar limit''). Surprisingly, the UAE is also appropriate to be employed to investigate behaviors in this limit~\cite{Prachi_concentricice2019}. Then the arguments above can be directly followed, except for the term $-r_{32}^{\rm H}(n,l;b)r_{21}^{\rm H}(n,l;a)$ in the instance above, which can be calculated, given $a\rightarrow\infty$, as
\begin{eqnarray}
& &
e^{-2\nu\eta(z_{2,b})+2\nu\eta(z_{2,a})}
\frac{
\varepsilon_2(1+z_{3,b}^2)^{1/2}
-
\varepsilon_3(1+z_{2,b}^2)^{1/2}
}{
\varepsilon_2(1+z_{3,b}^2)^{1/2}
+
\varepsilon_3(1+z_{2,b}^2)^{1/2}
}
\nonumber\\
& &\quad
\times
\frac{
\varepsilon_1(1+z_{2,a}^2)^{1/2}
-
\varepsilon_2(1+z_{1,a}^2)^{1/2}
}{
\varepsilon_1(1+z_{2,a}^2)^{1/2}
+
\varepsilon_1(1+z_{1,a}^2)^{1/2}
}
\nonumber\\
&\sim&
e^{-2\sqrt{\nu^2/a^2+\varepsilon_2\zeta_n^2}d}
\frac{
\varepsilon_2\sqrt{\nu^2/a^2+\varepsilon_3\zeta_n^2}
-
\varepsilon_3\sqrt{\nu^2/a^2+\varepsilon_2\zeta_n^2}
}{
\varepsilon_2\sqrt{\nu^2/a^2+\varepsilon_3\zeta_n^2}
+
\varepsilon_3\sqrt{\nu^2/a^2+\varepsilon_2\zeta_n^2}
}
\nonumber\\
&&
\times
\frac{
\varepsilon_1\sqrt{\nu^2/a^2+\varepsilon_2\zeta_n^2}
-
\varepsilon_2\sqrt{\nu^2/a^2+\varepsilon_1\zeta_n^2}
}{
\varepsilon_1\sqrt{\nu^2/a^2+\varepsilon_2\zeta_n^2}
+
\varepsilon_1\sqrt{\nu^2/a^2+\varepsilon_1\zeta_n^2}
}
.
\end{eqnarray}
By making the substitution $\nu/a\rightarrow k$ and correspondingly $\nu/a^2\rightarrow dkk$, the Casimir free energy per unit area can be expressed in terms of an integral over $k$ as
\begin{eqnarray}
\label{eq.FEalarge}
f_{\rm H,s}
&\sim&
T\psum_{n=0}^{\infty}\int_{0}^{\infty}\frac{dkk}{2\pi}\ln\bigg(
1
+
\frac{
\varepsilon_2\widetilde{\kappa}_{3,n}-\varepsilon_3\widetilde{\kappa}_{2,n}
}{
\varepsilon_2\widetilde{\kappa}_{3,n}+\varepsilon_3\widetilde{\kappa}_{2,n}
}
\nonumber\\
& &
\times
\frac{
\varepsilon_1\widetilde{\kappa}_{2,n}-\varepsilon_2\widetilde{\kappa}_{1,n}
}{
\varepsilon_1\widetilde{\kappa}_{2,n}+\varepsilon_2\widetilde{\kappa}_{1,n}
}e^{-2\widetilde{\kappa}_{2,n}d}
\bigg),
\end{eqnarray}
in which $\widetilde{\kappa}_{i,n}=\sqrt{k^2+\varepsilon_i\zeta_n^2}$, and the Casimir-Lifshitz free energy per unit area for the planar three-layer configuration is thus exactly derived.

\section*{Appendix: Casimir energy of multi-layer concentric configurations}
\label{App.MCC}

\par The Casimir energy of a concentric system is (The natural unit $\hbar=c=\varepsilon_0=\mu_0=k_B=1$ is used.)
\begin{eqnarray}
E^{\rm TE}
&=&
-\sum_{l=1}^{\infty}\nu\int\frac{d\zeta}{2\pi}\frac{1}{W_E}\bigg[
\frac{\mathfrak{e}'_+(r)}{\mu(r)}\frac{\partial\zeta\mathfrak{e}_-(r)}{\partial\zeta}
\nonumber\\
& &
-
\zeta\mathfrak{e}_+(r)\frac{\partial}{\partial\zeta}\frac{\mathfrak{e}'_-(r)}{\mu(r)}
\bigg]^{\infty}_{0},
\end{eqnarray}
and the TM contributions are obtained by making the EM-substitutions $\varepsilon\leftrightarrow\mu,\ \mathfrak{e}\leftrightarrow\mathfrak{h}$, which satisfy
\begin{eqnarray}
\bigg[\frac{d}{dr}\mu^{-1}\frac{d}{dr}-\frac{l(l+1)}{\mu r^2}-\varepsilon\zeta^2\bigg]
\mathfrak{e}_{\pm}(\zeta,l;r)&=&0,\\ \bigg[\frac{d}{dr}\varepsilon^{-1}\frac{d}{dr}-\frac{l(l+1)}{\varepsilon r^2}-\mu\zeta^2\bigg]
\mathfrak{h}_{\pm}(\zeta,l;r)&=&0,
\end{eqnarray}
with $+$ and $-$ signifying the solution being $0$ and finite at the infinity and origin, respectively. Within a uniform background, the solutions are denoted as $\mathfrak{f}_{\pm}(\zeta,l; r)=\mathfrak{e}_{\pm}(\zeta,l;\kappa r)=\mathfrak{h}_{\pm}(\zeta,l;\kappa r)$, and $\kappa=\sqrt{\varepsilon\mu}|\zeta|$. Then, for a layered concentric system with the permittivity and permeability being
\begin{eqnarray}
(\varepsilon,\mu)(r)=
\left\{
  \begin{array}{cc}
    (\varepsilon_1,\mu_1), & r<a_1, \\
    (\varepsilon_2,\mu_2), & a_1<r<a_2, \\
    \vdots & \vdots \\
    (\varepsilon_{N},\mu_{N}), & a_{N-1}<r<a_N, \\
    (\varepsilon_{N+1},\mu_{N+1}), & r>a_N, \\
  \end{array}
\right.
\end{eqnarray}
the solutions, for instance $\mathfrak{e}_\pm$, are
\begin{widetext}
\begin{eqnarray}
\mathfrak{e}_+(r)=
\left\{
  \begin{array}{cc}
    A_1\mathfrak{f}_{1,+}(r)+B_1\mathfrak{f}_{1,-}(r), & r<a_1, \\
    A_2\mathfrak{f}_{2,+}(r)+B_2\mathfrak{f}_{2,-}(r), & a_1<r<a_2, \\
    \vdots & \vdots \\
    A_N\mathfrak{f}_{N,+}(r)+B_N\mathfrak{f}_{N,-}(r), & a_{N-1}<r<a_N, \\
    \mathfrak{f}_{N+1,+}(r), & r>a_N, \\
  \end{array}
\right.
\mathfrak{e}_-(r)=
\left\{
  \begin{array}{cc}
    \mathfrak{f}_{1,-}(r), & r<a_1, \\
    C_2\mathfrak{f}_{2,+}(r)+D_2\mathfrak{f}_{2,-}(r), & a_1<r<a_2, \\
    \vdots & \vdots \\
    C_N\mathfrak{f}_{N,+}(r)+D_N\mathfrak{f}_{N,-}(r), & a_{N-1}<r<a_N, \\
    C_{N+1}\mathfrak{f}_{N+1,+}(r)+D_{N+1}\mathfrak{f}_{N+1,-}(r), & r>a_N, \\
  \end{array}
\right.
\end{eqnarray}
The matching conditions leads us to the recursive expressions for those coefficients, taking $\mathfrak{e}_{\pm}$ as the instance without losing any generity, above as
\begin{subequations}
\begin{eqnarray}
A_{i}=A_{i+1}\frac{[\mathfrak{f}_{i+1,+},\mathfrak{f}_{i,-}]_{\mu}(a_i)}{W_{i}}
+
B_{i+1}\frac{[\mathfrak{f}_{i+1,-},\mathfrak{f}_{i,-}]_{\mu}(a_i)}{W_{i}},\
B_{i}=A_{i+1}\frac{[\mathfrak{f}_{i,+},\mathfrak{f}_{i+1,+}]_{\mu}(a_i)}{W_{i}}
+
B_{i+1}\frac{[\mathfrak{f}_{i,+},\mathfrak{f}_{i+1,-}]_{\mu}(a_i)}{W_{i}},\quad
\end{eqnarray}
\begin{eqnarray}
A_N=\frac{[\mathfrak{f}_{N+1,+},\mathfrak{f}_{N,-}]_{\mu}(a_N)}{W_N},\ B_N=\frac{[\mathfrak{f}_{N,+},\mathfrak{f}_{N+1,+}]_{\mu}(a_N)}{W_N},
\end{eqnarray}
\begin{eqnarray}
C_{i+1}=C_i\frac{[\mathfrak{f}_{i,+},\mathfrak{f}_{i+1,-}]_{\mu}(a_i)}{W_{i+1}}
+
D_i\frac{[\mathfrak{f}_{i,-},\mathfrak{f}_{i+1,-}]_{\mu}(a_i)}{W_{i+1}},\
D_{i+1}=C_i\frac{[\mathfrak{f}_{i+1,+},\mathfrak{f}_{i,+}]_{\mu}(a_i)}{W_{i+1}}
+
D_i\frac{[\mathfrak{f}_{i+1,+},\mathfrak{f}_{i,-}]_{\mu}(a_i)}{W_{i+1}},
\end{eqnarray}
\begin{eqnarray}
C_2=\frac{[\mathfrak{f}_{1,-},\mathfrak{f}_{2,-}]_{\mu}(a_1)}{W_2},\ D_2=\frac{[\mathfrak{f}_{2,+},\mathfrak{f}_{1,-}]_{\mu}(a_1)}{W_2}.
\end{eqnarray}
\end{subequations}
For example, $N=2$ case can be written as
\begin{subequations}
\begin{eqnarray}
A_{1}=\frac{[\mathfrak{f}_{3,+},\mathfrak{f}_{2,-}]_{\mu}(a_2)}{W_2}
\frac{[\mathfrak{f}_{2,+},\mathfrak{f}_{1,-}]_{\mu}(a_1)}{W_{1}}
+
\frac{[\mathfrak{f}_{2,+},\mathfrak{f}_{3,+}]_{\mu}(a_2)}{W_2}
\frac{[\mathfrak{f}_{2,-},\mathfrak{f}_{1,-}]_{\mu}(a_1)}{W_{1}},
\end{eqnarray}
\begin{eqnarray}
A_2=\frac{[\mathfrak{f}_{3,+},\mathfrak{f}_{2,-}]_{\mu}(a_2)}{W_2},\ B_2=\frac{[\mathfrak{f}_{2,+},\mathfrak{f}_{3,+}]_{\mu}(a_2)}{W_2},
\end{eqnarray}
\begin{eqnarray}
D_{3}=\frac{[\mathfrak{f}_{1,-},\mathfrak{f}_{2,-}]_{\mu}(a_1)}{W_2}
\frac{[\mathfrak{f}_{3,+},\mathfrak{f}_{2,+}]_{\mu}(a_2)}{W_{3}}
+
\frac{[\mathfrak{f}_{2,+},\mathfrak{f}_{1,-}]_{\mu}(a_1)}{W_2}
\frac{[\mathfrak{f}_{3,+},\mathfrak{f}_{2,-}]_{\mu}(a_2)}{W_{3}},
\end{eqnarray}
\begin{eqnarray}
C_2=\frac{[\mathfrak{f}_{1,-},\mathfrak{f}_{2,-}]_{\mu}(a_1)}{W_2},\ D_2=\frac{[\mathfrak{f}_{2,+},\mathfrak{f}_{1,-}]_{\mu}(a_1)}{W_2},
\end{eqnarray}
\begin{eqnarray}
E^{\rm TE}=\sum_{l=1}^{\infty}\nu\int\frac{d\zeta}{2\pi}\ln D_3
=\sum_{l=1}^{\infty}\nu\int\frac{d\zeta}{2\pi}\ln\bigg\{1-
\frac{
[\mathfrak{f}_{3,+},\mathfrak{f}_{2,+}]_{\mu}(a_2)[\mathfrak{f}_{2,-},\mathfrak{f}_{1,-}]_{\mu}(a_1)
}{
[\mathfrak{f}_{3,+},\mathfrak{f}_{2,-}]_{\mu}(a_2)[\mathfrak{f}_{2,+},\mathfrak{f}_{1,-}]_{\mu}(a_1)
}
\bigg\}
,
\end{eqnarray}
\end{subequations}
where the bulk terms are subtracted in the first step, and the surface divergence is subtracted in the second step. To find $D_{N+1}$, we write
\begin{subequations}
\begin{eqnarray}
\left[
  \begin{array}{c}
    C_{i+1} \\
    D_{i+1} \\
  \end{array}
\right]=
M_{i+1,i}
\left[
  \begin{array}{c}
    C_i \\
    D_i \\
  \end{array}
\right]
=
\left[
  \begin{array}{cc}
    \frac{[\mathfrak{f}_{i,+},\mathfrak{f}_{i+1,-}]_{\mu}(a_i)}{W_{i+1}} & \frac{[\mathfrak{f}_{i,-},\mathfrak{f}_{i+1,-}]_{\mu}(a_i)}{W_{i+1}} \\
    \frac{[\mathfrak{f}_{i+1,+},\mathfrak{f}_{i,+}]_{\mu}(a_i)}{W_{i+1}} & \frac{[\mathfrak{f}_{i+1,+},\mathfrak{f}_{i,-}]_{\mu}(a_i)}{W_{i+1}} \\
  \end{array}
\right]
\left[
  \begin{array}{c}
    C_i \\
    D_i \\
  \end{array}
\right],
\end{eqnarray}
which gives us
\begin{eqnarray}
E^{\rm TE}=\sum_{l=1}^{\infty}\nu\int\frac{d\zeta}{2\pi}\ln D_{N+1}
=\sum_{l=1}^{\infty}\nu\int\frac{d\zeta}{2\pi}\ln\bigg\{1+
\frac{
M_{N+1,N}^{2,j_N}\cdots M_{i+1,i}^{j_{i+1},j_i}\cdots M_{2,1}^{j_2,2}
}{
M_{N+1,N}^{2,2}\cdots M_{i+1,i}^{2,2}\cdots M_{2,1}^{2,2}
}
\bigg\}
.
\end{eqnarray}
\end{subequations}
The corresponding TM contribution can be obtained in the same way.
\end{widetext}

\begin{acknowledgments}
We thank the Terahertz Physics and Devices Group, Nanchang
University for the strong computational facility support.
Y.L. was supported by the National Natural Science Foundation
of China (Grant No. 12304396). P.P. thanks T. Jayasekara
for collaborative assistance during the course of this work.
The research by O.I.M. and M.B. is part of Project No.
2022/47/P/ST3/01236 cofunded by the National Science
Centre and the European Union's Horizon 2020 research and
innovation program under the Marie Sk?odowska-Curie Grant
Agreement No. 945339.
Institutional and infrastructural support for the ENSEMBLE3 Centre of
Excellence was provided through the ENSEMBLE3 project (MAB/2020/14)
delivered within the Foundation for Polish Science International
Research Agenda Programme and co-financed by the European Regional
Development Fund and the Horizon 2020 Teaming for Excellence initiative
(Grant Agreement No. 857543), as well as the Ministry of Education and
Science initiative ``Support for Centres of Excellence in Poland under
Horizon 2020''.
We gratefully acknowledge Poland's high-performance computing
infrastructure PLGrid (HPC Centers: ACK Cyfronet
AGH) for providing computer facilities and support within
computational Grant No. PLG/2023/016228. C.P. thanks the
European Union¡¯s Horizon 2020 research and innovation program
(GA No. 101058694).

\end{acknowledgments}

\bibliography{PRB_drywet_bib}

\begin{thebibliography}{41}%
\makeatletter
\providecommand \@ifxundefined [1]{%
 \@ifx{#1\undefined}
}%
\providecommand \@ifnum [1]{%
 \ifnum #1\expandafter \@firstoftwo
 \else \expandafter \@secondoftwo
 \fi
}%
\providecommand \@ifx [1]{%
 \ifx #1\expandafter \@firstoftwo
 \else \expandafter \@secondoftwo
 \fi
}%
\providecommand \natexlab [1]{#1}%
\providecommand \enquote  [1]{``#1''}%
\providecommand \bibnamefont  [1]{#1}%
\providecommand \bibfnamefont [1]{#1}%
\providecommand \citenamefont [1]{#1}%
\providecommand \href@noop [0]{\@secondoftwo}%
\providecommand \href [0]{\begingroup \@sanitize@url \@href}%
\providecommand \@href[1]{\@@startlink{#1}\@@href}%
\providecommand \@@href[1]{\endgroup#1\@@endlink}%
\providecommand \@sanitize@url [0]{\catcode `\\12\catcode `\$12\catcode
  `\&12\catcode `\#12\catcode `\^12\catcode `\_12\catcode `\%12\relax}%
\providecommand \@@startlink[1]{}%
\providecommand \@@endlink[0]{}%
\providecommand \url  [0]{\begingroup\@sanitize@url \@url }%
\providecommand \@url [1]{\endgroup\@href {#1}{\urlprefix }}%
\providecommand \urlprefix  [0]{URL }%
\providecommand \Eprint [0]{\href }%
\providecommand \doibase [0]{http://dx.doi.org/}%
\providecommand \selectlanguage [0]{\@gobble}%
\providecommand \bibinfo  [0]{\@secondoftwo}%
\providecommand \bibfield  [0]{\@secondoftwo}%
\providecommand \translation [1]{[#1]}%
\providecommand \BibitemOpen [0]{}%
\providecommand \bibitemStop [0]{}%
\providecommand \bibitemNoStop [0]{.\EOS\space}%
\providecommand \EOS [0]{\spacefactor3000\relax}%
\providecommand \BibitemShut  [1]{\csname bibitem#1\endcsname}%
\let\auto@bib@innerbib\@empty
\bibitem [{\citenamefont {M\"ulmenst\"adt}\ \emph {et~al.}(2015)\citenamefont
  {M\"ulmenst\"adt}, \citenamefont {Sourdeval}, \citenamefont {Delano\"e},\
  and\ \citenamefont {Quaas}}]{MulenstadtSourdevalDalanoeQuaasGRL2015}%
  \BibitemOpen
  \bibfield  {author} {\bibinfo {author} {\bibfnamefont {J.}~\bibnamefont
  {M\"ulmenst\"adt}}, \bibinfo {author} {\bibfnamefont {O.}~\bibnamefont
  {Sourdeval}}, \bibinfo {author} {\bibfnamefont {J.}~\bibnamefont
  {Delano\"e}}, \ and\ \bibinfo {author} {\bibfnamefont {J.}~\bibnamefont
  {Quaas}},\ }\bibfield  {title} {\enquote {\bibinfo {title} {Frequency of
  occurrence of rain from liquid-, mixed-, and ice-phase clouds derived from
  a-train satellite retrievals},}\ }\href {\doibase
  https://doi.org/10.1002/2015GL064604} {\bibfield  {journal} {\bibinfo
  {journal} {Geophys. Res. Lett.}\ }\textbf {\bibinfo {volume} {42}},\ \bibinfo
  {pages} {6502--6509} (\bibinfo {year} {2015})}\BibitemShut {NoStop}%
\bibitem [{\citenamefont {Storelvmo}\ and\ \citenamefont
  {Tan}(2015)}]{StorelvmoTan2015}%
  \BibitemOpen
  \bibfield  {author} {\bibinfo {author} {\bibfnamefont {T.}~\bibnamefont
  {Storelvmo}}\ and\ \bibinfo {author} {\bibfnamefont {I.}~\bibnamefont
  {Tan}},\ }\bibfield  {title} {\enquote {\bibinfo {title} {{The
  Wegener-Bergeron-Findeisen process - Its discovery and vital importance for
  weather and climate}},}\ }\href@noop {} {\bibfield  {journal} {\bibinfo
  {journal} {Meteorol. Zeitschrift}\ }\textbf {\bibinfo {volume} {24}},\
  \bibinfo {pages} {455--461} (\bibinfo {year} {2015})}\BibitemShut {NoStop}%
\bibitem [{\citenamefont {Arienti}\ \emph {et~al.}(2018)\citenamefont
  {Arienti}, \citenamefont {Geier}, \citenamefont {Yang}, \citenamefont
  {Orcutt}, \citenamefont {Zenker},\ and\ \citenamefont
  {Brooks}}]{arienti2018experimental}%
  \BibitemOpen
  \bibfield  {author} {\bibinfo {author} {\bibfnamefont {M.}~\bibnamefont
  {Arienti}}, \bibinfo {author} {\bibfnamefont {M.}~\bibnamefont {Geier}},
  \bibinfo {author} {\bibfnamefont {X.}~\bibnamefont {Yang}}, \bibinfo {author}
  {\bibfnamefont {J.}~\bibnamefont {Orcutt}}, \bibinfo {author} {\bibfnamefont
  {J.}~\bibnamefont {Zenker}}, \ and\ \bibinfo {author} {\bibfnamefont {S.~D.}\
  \bibnamefont {Brooks}},\ }\bibfield  {title} {\enquote {\bibinfo {title} {An
  experimental and numerical study of the light scattering properties of ice
  crystals with black carbon inclusions},}\ }\href@noop {} {\bibfield
  {journal} {\bibinfo  {journal} {J. Quant. Spectrosc. Radiat. Transf.}\
  }\textbf {\bibinfo {volume} {211}},\ \bibinfo {pages} {50--63} (\bibinfo
  {year} {2018})}\BibitemShut {NoStop}%
\bibitem [{\citenamefont {Zeng}\ \emph {et~al.}(2009)\citenamefont {Zeng},
  \citenamefont {Tao}, \citenamefont {Zhang}, \citenamefont {Hou},
  \citenamefont {Xie}, \citenamefont {Lang}, \citenamefont {Li}, \citenamefont
  {Starr}, \citenamefont {Li},\ and\ \citenamefont
  {Simpson}}]{zeng2009indirect}%
  \BibitemOpen
  \bibfield  {author} {\bibinfo {author} {\bibfnamefont {X.}~\bibnamefont
  {Zeng}}, \bibinfo {author} {\bibfnamefont {W.~K.}\ \bibnamefont {Tao}},
  \bibinfo {author} {\bibfnamefont {M.}~\bibnamefont {Zhang}}, \bibinfo
  {author} {\bibfnamefont {A.~Y.}\ \bibnamefont {Hou}}, \bibinfo {author}
  {\bibfnamefont {S.}~\bibnamefont {Xie}}, \bibinfo {author} {\bibfnamefont
  {S.}~\bibnamefont {Lang}}, \bibinfo {author} {\bibfnamefont {X.}~\bibnamefont
  {Li}}, \bibinfo {author} {\bibfnamefont {D.~O'C}\ \bibnamefont {Starr}},
  \bibinfo {author} {\bibfnamefont {X.}~\bibnamefont {Li}}, \ and\ \bibinfo
  {author} {\bibfnamefont {J.}~\bibnamefont {Simpson}},\ }\bibfield  {title}
  {\enquote {\bibinfo {title} {An indirect effect of ice nuclei on atmospheric
  radiation},}\ }\href@noop {} {\bibfield  {journal} {\bibinfo  {journal} {J.
  Atmos. Sci.}\ }\textbf {\bibinfo {volume} {66}},\ \bibinfo {pages} {41--61}
  (\bibinfo {year} {2009})}\BibitemShut {NoStop}%
\bibitem [{\citenamefont {Hawker}\ \emph {et~al.}(2021)\citenamefont {Hawker},
  \citenamefont {Miltenberger}, \citenamefont {Wilkinson}, \citenamefont
  {Hill}, \citenamefont {Shipway}, \citenamefont {Cui}, \citenamefont {Cotton},
  \citenamefont {Carslaw}, \citenamefont {Field},\ and\ \citenamefont
  {Murray}}]{hawker2021temperature}%
  \BibitemOpen
  \bibfield  {author} {\bibinfo {author} {\bibfnamefont {R.~E.}\ \bibnamefont
  {Hawker}}, \bibinfo {author} {\bibfnamefont {A.~K.}\ \bibnamefont
  {Miltenberger}}, \bibinfo {author} {\bibfnamefont {J.~M.}\ \bibnamefont
  {Wilkinson}}, \bibinfo {author} {\bibfnamefont {A.~A.}\ \bibnamefont {Hill}},
  \bibinfo {author} {\bibfnamefont {B.~J.}\ \bibnamefont {Shipway}}, \bibinfo
  {author} {\bibfnamefont {Z.}~\bibnamefont {Cui}}, \bibinfo {author}
  {\bibfnamefont {R.~J.}\ \bibnamefont {Cotton}}, \bibinfo {author}
  {\bibfnamefont {K.~S.}\ \bibnamefont {Carslaw}}, \bibinfo {author}
  {\bibfnamefont {P.~R.}\ \bibnamefont {Field}}, \ and\ \bibinfo {author}
  {\bibfnamefont {B.~J.}\ \bibnamefont {Murray}},\ }\bibfield  {title}
  {\enquote {\bibinfo {title} {The temperature dependence of ice-nucleating
  particle concentrations affects the radiative properties of tropical
  convective cloud systems},}\ }\href@noop {} {\bibfield  {journal} {\bibinfo
  {journal} {Atmos. Chem. Phys.}\ }\textbf {\bibinfo {volume} {21}},\ \bibinfo
  {pages} {5439--5461} (\bibinfo {year} {2021})}\BibitemShut {NoStop}%
\bibitem [{\citenamefont {Burrows}\ \emph {et~al.}(2022)\citenamefont
  {Burrows}, \citenamefont {McCluskey}, \citenamefont {Cornwell}, \citenamefont
  {Steinke}, \citenamefont {Zhang}, \citenamefont {Zhao}, \citenamefont
  {Zawadowicz}, \citenamefont {Raman}, \citenamefont {Kulkarni}, \citenamefont
  {China}, \citenamefont {Zelenyuk},\ and\ \citenamefont
  {DeMott}}]{burrows2022ice}%
  \BibitemOpen
  \bibfield  {author} {\bibinfo {author} {\bibfnamefont {S.~M.}\ \bibnamefont
  {Burrows}}, \bibinfo {author} {\bibfnamefont {C.~S.}\ \bibnamefont
  {McCluskey}}, \bibinfo {author} {\bibfnamefont {G.}~\bibnamefont {Cornwell}},
  \bibinfo {author} {\bibfnamefont {I.}~\bibnamefont {Steinke}}, \bibinfo
  {author} {\bibfnamefont {K.}~\bibnamefont {Zhang}}, \bibinfo {author}
  {\bibfnamefont {B.}~\bibnamefont {Zhao}}, \bibinfo {author} {\bibfnamefont
  {M.}~\bibnamefont {Zawadowicz}}, \bibinfo {author} {\bibfnamefont
  {A.}~\bibnamefont {Raman}}, \bibinfo {author} {\bibfnamefont
  {G.}~\bibnamefont {Kulkarni}}, \bibinfo {author} {\bibfnamefont
  {S.}~\bibnamefont {China}}, \bibinfo {author} {\bibfnamefont
  {A.}~\bibnamefont {Zelenyuk}}, \ and\ \bibinfo {author} {\bibfnamefont
  {P.~J.}\ \bibnamefont {DeMott}},\ }\bibfield  {title} {\enquote {\bibinfo
  {title} {{Ice-Nucleating Particles That Impact Clouds and Climate:
  Observational and Modeling Research Needs}},}\ }\href@noop {} {\bibfield
  {journal} {\bibinfo  {journal} {Rev. Geophys.}\ }\textbf {\bibinfo {volume}
  {60}},\ \bibinfo {pages} {e2021RG000745} (\bibinfo {year}
  {2022})}\BibitemShut {NoStop}%
\bibitem [{\citenamefont {Maeda}(2021)}]{maeda2021brief}%
  \BibitemOpen
  \bibfield  {author} {\bibinfo {author} {\bibfnamefont {N.}~\bibnamefont
  {Maeda}},\ }\bibfield  {title} {\enquote {\bibinfo {title} {Brief overview of
  ice nucleation},}\ }\href@noop {} {\bibfield  {journal} {\bibinfo  {journal}
  {Molecules}\ }\textbf {\bibinfo {volume} {26}},\ \bibinfo {pages} {392}
  (\bibinfo {year} {2021})}\BibitemShut {NoStop}%
\bibitem [{\citenamefont {Findeisen}\ \emph {et~al.}(2015)\citenamefont
  {Findeisen}, \citenamefont {Volken}, \citenamefont {Giesche},\ and\
  \citenamefont {Br{\"o}nnimann}}]{findeisen2015}%
  \BibitemOpen
  \bibfield  {author} {\bibinfo {author} {\bibfnamefont {W.}~\bibnamefont
  {Findeisen}}, \bibinfo {author} {\bibfnamefont {E.}~\bibnamefont {Volken}},
  \bibinfo {author} {\bibfnamefont {A.~M.}\ \bibnamefont {Giesche}}, \ and\
  \bibinfo {author} {\bibfnamefont {S.}~\bibnamefont {Br{\"o}nnimann}},\
  }\bibfield  {title} {\enquote {\bibinfo {title} {Colloidal meteorological
  processes in the formation of precipitation},}\ }\href {\doibase
  10.1127/metz/2015/0675} {\bibfield  {journal} {\bibinfo  {journal} {Meteorol.
  Zeitschrift}\ }\textbf {\bibinfo {volume} {24}},\ \bibinfo {pages} {443--454}
  (\bibinfo {year} {2015})}\BibitemShut {NoStop}%
\bibitem [{\citenamefont {Vali}(1966)}]{ValiIceNucleai_Nature1966}%
  \BibitemOpen
  \bibfield  {author} {\bibinfo {author} {\bibfnamefont {G.}~\bibnamefont
  {Vali}},\ }\bibfield  {title} {\enquote {\bibinfo {title} {Sizes of
  atmospheric ice nuclei},}\ }\href {\doibase 10.1038/212384a0} {\bibfield
  {journal} {\bibinfo  {journal} {Nature}\ }\textbf {\bibinfo {volume} {212}},\
  \bibinfo {pages} {384--385} (\bibinfo {year} {1966})}\BibitemShut {NoStop}%
\bibitem [{\citenamefont {Atkinson}\ \emph {et~al.}(2013)\citenamefont
  {Atkinson}, \citenamefont {Murray}, \citenamefont {Woodhouse}, \citenamefont
  {Whale}, \citenamefont {Baustian}, \citenamefont {Carslaw}, \citenamefont
  {Dobbie}, \citenamefont {O'Sullivan},\ and\ \citenamefont
  {Malkin}}]{atkinson2013importance}%
  \BibitemOpen
  \bibfield  {author} {\bibinfo {author} {\bibfnamefont {J.~D.}\ \bibnamefont
  {Atkinson}}, \bibinfo {author} {\bibfnamefont {B.~J.}\ \bibnamefont
  {Murray}}, \bibinfo {author} {\bibfnamefont {M.~T.}\ \bibnamefont
  {Woodhouse}}, \bibinfo {author} {\bibfnamefont {T.~F.}\ \bibnamefont
  {Whale}}, \bibinfo {author} {\bibfnamefont {K.~J.}\ \bibnamefont {Baustian}},
  \bibinfo {author} {\bibfnamefont {K.~S.}\ \bibnamefont {Carslaw}}, \bibinfo
  {author} {\bibfnamefont {S.}~\bibnamefont {Dobbie}}, \bibinfo {author}
  {\bibfnamefont {D.}~\bibnamefont {O'Sullivan}}, \ and\ \bibinfo {author}
  {\bibfnamefont {T.~L.}\ \bibnamefont {Malkin}},\ }\bibfield  {title}
  {\enquote {\bibinfo {title} {The importance of feldspar for ice nucleation by
  mineral dust in mixed-phase clouds},}\ }\href {\doibase 10.1038/nature12278}
  {\bibfield  {journal} {\bibinfo  {journal} {Nature}\ }\textbf {\bibinfo
  {volume} {498}},\ \bibinfo {pages} {355--358} (\bibinfo {year}
  {2013})}\BibitemShut {NoStop}%
\bibitem [{\citenamefont {Murray}(2017)}]{Murraydoi:10.1126/science.aam5320}%
  \BibitemOpen
  \bibfield  {author} {\bibinfo {author} {\bibfnamefont {B.~J.}\ \bibnamefont
  {Murray}},\ }\bibfield  {title} {\enquote {\bibinfo {title} {Cracking the
  problem of ice nucleation},}\ }\href {\doibase 10.1126/science.aam5320}
  {\bibfield  {journal} {\bibinfo  {journal} {Science}\ }\textbf {\bibinfo
  {volume} {355}},\ \bibinfo {pages} {346--347} (\bibinfo {year}
  {2017})}\BibitemShut {NoStop}%
\bibitem [{\citenamefont {Kiselev}\ \emph {et~al.}(2017)\citenamefont
  {Kiselev}, \citenamefont {Bachmann}, \citenamefont {Pedevilla}, \citenamefont
  {Cox}, \citenamefont {Michaelides}, \citenamefont {Gerthsen},\ and\
  \citenamefont {Leisner}}]{kiselev2017active}%
  \BibitemOpen
  \bibfield  {author} {\bibinfo {author} {\bibfnamefont {A.}~\bibnamefont
  {Kiselev}}, \bibinfo {author} {\bibfnamefont {F.}~\bibnamefont {Bachmann}},
  \bibinfo {author} {\bibfnamefont {P.}~\bibnamefont {Pedevilla}}, \bibinfo
  {author} {\bibfnamefont {S.~J.}\ \bibnamefont {Cox}}, \bibinfo {author}
  {\bibfnamefont {A.}~\bibnamefont {Michaelides}}, \bibinfo {author}
  {\bibfnamefont {D.}~\bibnamefont {Gerthsen}}, \ and\ \bibinfo {author}
  {\bibfnamefont {T.}~\bibnamefont {Leisner}},\ }\bibfield  {title} {\enquote
  {\bibinfo {title} {Active sites in heterogeneous ice nucleation-the example
  of k-rich feldspars},}\ }\href {\doibase 10.1126/science.aai8034} {\bibfield
  {journal} {\bibinfo  {journal} {Science}\ }\textbf {\bibinfo {volume}
  {355}},\ \bibinfo {pages} {367--371} (\bibinfo {year} {2017})}\BibitemShut
  {NoStop}%
\bibitem [{\citenamefont {Lloyd}\ \emph {et~al.}(2020)\citenamefont {Lloyd},
  \citenamefont {Choularton}, \citenamefont {Bower}, \citenamefont {Crosier},
  \citenamefont {Gallagher}, \citenamefont {Flynn}, \citenamefont {Dorsey},
  \citenamefont {Liu}, \citenamefont {Taylor}, \citenamefont {Schlenczek},
  \citenamefont {Fugal}, \citenamefont {Borrmann}, \citenamefont {Cotton},
  \citenamefont {Field},\ and\ \citenamefont
  {Blyth}}]{Lloydetalacp-20-3895-2020}%
  \BibitemOpen
  \bibfield  {author} {\bibinfo {author} {\bibfnamefont {G.}~\bibnamefont
  {Lloyd}}, \bibinfo {author} {\bibfnamefont {T.}~\bibnamefont {Choularton}},
  \bibinfo {author} {\bibfnamefont {K.}~\bibnamefont {Bower}}, \bibinfo
  {author} {\bibfnamefont {J.}~\bibnamefont {Crosier}}, \bibinfo {author}
  {\bibfnamefont {M.}~\bibnamefont {Gallagher}}, \bibinfo {author}
  {\bibfnamefont {M.}~\bibnamefont {Flynn}}, \bibinfo {author} {\bibfnamefont
  {J.}~\bibnamefont {Dorsey}}, \bibinfo {author} {\bibfnamefont
  {D.}~\bibnamefont {Liu}}, \bibinfo {author} {\bibfnamefont {J.~W.}\
  \bibnamefont {Taylor}}, \bibinfo {author} {\bibfnamefont {O.}~\bibnamefont
  {Schlenczek}}, \bibinfo {author} {\bibfnamefont {J.}~\bibnamefont {Fugal}},
  \bibinfo {author} {\bibfnamefont {S.}~\bibnamefont {Borrmann}}, \bibinfo
  {author} {\bibfnamefont {R.}~\bibnamefont {Cotton}}, \bibinfo {author}
  {\bibfnamefont {P.}~\bibnamefont {Field}}, \ and\ \bibinfo {author}
  {\bibfnamefont {A.}~\bibnamefont {Blyth}},\ }\bibfield  {title} {\enquote
  {\bibinfo {title} {Small ice particles at slightly supercooled temperatures
  in tropical maritime convection},}\ }\href {\doibase
  10.5194/acp-20-3895-2020} {\bibfield  {journal} {\bibinfo  {journal} {Atmos.
  Chem. Phys.}\ }\textbf {\bibinfo {volume} {20}},\ \bibinfo {pages}
  {3895--3904} (\bibinfo {year} {2020})}\BibitemShut {NoStop}%
\bibitem [{\citenamefont {Aragones}\ \emph {et~al.}(2011)\citenamefont
  {Aragones}, \citenamefont {MacDowell},\ and\ \citenamefont
  {Vega}}]{AragonesMacDowellVega2011}%
  \BibitemOpen
  \bibfield  {author} {\bibinfo {author} {\bibfnamefont {J.~L.}\ \bibnamefont
  {Aragones}}, \bibinfo {author} {\bibfnamefont {L.~G.}\ \bibnamefont
  {MacDowell}}, \ and\ \bibinfo {author} {\bibfnamefont {C.}~\bibnamefont
  {Vega}},\ }\bibfield  {title} {\enquote {\bibinfo {title} {Dielectric
  constant of ices and water: A lesson about water interactions},}\ }\href
  {\doibase 10.1021/jp105975c} {\bibfield  {journal} {\bibinfo  {journal} {J.
  Phys. Chem. A}\ }\textbf {\bibinfo {volume} {115}},\ \bibinfo {pages}
  {5745--5758} (\bibinfo {year} {2011})}\BibitemShut {NoStop}%
\bibitem [{\citenamefont {Fiedler}\ \emph {et~al.}(2020)\citenamefont
  {Fiedler}, \citenamefont {Bostr\"om}, \citenamefont {Persson}, \citenamefont
  {Brevik}, \citenamefont {Corkery}, \citenamefont {Buhmann},\ and\
  \citenamefont {Parsons}}]{JohannesWater2019}%
  \BibitemOpen
  \bibfield  {author} {\bibinfo {author} {\bibfnamefont {J.}~\bibnamefont
  {Fiedler}}, \bibinfo {author} {\bibfnamefont {M.}~\bibnamefont {Bostr\"om}},
  \bibinfo {author} {\bibfnamefont {C.}~\bibnamefont {Persson}}, \bibinfo
  {author} {\bibfnamefont {I.~H.}\ \bibnamefont {Brevik}}, \bibinfo {author}
  {\bibfnamefont {R.~W.}\ \bibnamefont {Corkery}}, \bibinfo {author}
  {\bibfnamefont {S.~Y.}\ \bibnamefont {Buhmann}}, \ and\ \bibinfo {author}
  {\bibfnamefont {D.~F.}\ \bibnamefont {Parsons}},\ }\bibfield  {title}
  {\enquote {\bibinfo {title} {Full-spectrum high resolution modeling of the
  dielectric function of water},}\ }\href {\doibase 10.1021/acs.jpcb.0c00410}
  {\bibfield  {journal} {\bibinfo  {journal} {J.\ Phys.\ Chem.\ B}\ }\textbf
  {\bibinfo {volume} {124}},\ \bibinfo {pages} {3103--3113} (\bibinfo {year}
  {2020})}\BibitemShut {NoStop}%
\bibitem [{\citenamefont {Luengo-M\'arquez}\ and\ \citenamefont
  {MacDowell}(2021)}]{LUENGOMARQUEZMacDowell2021}%
  \BibitemOpen
  \bibfield  {author} {\bibinfo {author} {\bibfnamefont {J.}~\bibnamefont
  {Luengo-M\'arquez}}\ and\ \bibinfo {author} {\bibfnamefont {L.~G.}\
  \bibnamefont {MacDowell}},\ }\bibfield  {title} {\enquote {\bibinfo {title}
  {Lifshitz theory of wetting films at three phase coexistence: The case of ice
  nucleation on silver iodide (agi)},}\ }\href {\doibase
  https://doi.org/10.1016/j.jcis.2021.01.060} {\bibfield  {journal} {\bibinfo
  {journal} {J. Coll. Interf. Sci.}\ }\textbf {\bibinfo {volume} {590}},\
  \bibinfo {pages} {527--538} (\bibinfo {year} {2021})}\BibitemShut {NoStop}%
\bibitem [{\citenamefont {Luengo-Marquez}\ \emph {et~al.}(2022)\citenamefont
  {Luengo-Marquez}, \citenamefont {Izquierdo-Ruiz},\ and\ \citenamefont
  {MacDowell}}]{LuengoMarquez_IzquierdoRuiz_MacDowell2022}%
  \BibitemOpen
  \bibfield  {author} {\bibinfo {author} {\bibfnamefont {J.}~\bibnamefont
  {Luengo-Marquez}}, \bibinfo {author} {\bibfnamefont {F.}~\bibnamefont
  {Izquierdo-Ruiz}}, \ and\ \bibinfo {author} {\bibfnamefont {L.~G.}\
  \bibnamefont {MacDowell}},\ }\bibfield  {title} {\enquote {\bibinfo {title}
  {Intermolecular forces at ice and water interfaces: Premelting, surface
  freezing, and regelation},}\ }\href {\doibase 10.1063/5.0097378} {\bibfield
  {journal} {\bibinfo  {journal} {J. Chem. Phys.}\ }\textbf {\bibinfo {volume}
  {157}},\ \bibinfo {pages} {044704} (\bibinfo {year} {2022})}\BibitemShut
  {NoStop}%
\bibitem [{\citenamefont {Elbaum}\ and\ \citenamefont
  {Schick}(1991{\natexlab{a}})}]{Elbaum2}%
  \BibitemOpen
  \bibfield  {author} {\bibinfo {author} {\bibfnamefont {M.}~\bibnamefont
  {Elbaum}}\ and\ \bibinfo {author} {\bibfnamefont {M.}~\bibnamefont
  {Schick}},\ }\bibfield  {title} {\enquote {\bibinfo {title} {{On the failure
  of water to freeze from its surface}},}\ }\href {\doibase
  10.1051/jp1:1991233} {\bibfield  {journal} {\bibinfo  {journal} {J. Phys. I
  France}\ }\textbf {\bibinfo {volume} {1}},\ \bibinfo {pages} {1665--1668}
  (\bibinfo {year} {1991}{\natexlab{a}})}\BibitemShut {NoStop}%
\bibitem [{\citenamefont {Elbaum}\ and\ \citenamefont
  {Schick}(1991{\natexlab{b}})}]{Elbaum}%
  \BibitemOpen
  \bibfield  {author} {\bibinfo {author} {\bibfnamefont {M.}~\bibnamefont
  {Elbaum}}\ and\ \bibinfo {author} {\bibfnamefont {M.}~\bibnamefont
  {Schick}},\ }\bibfield  {title} {\enquote {\bibinfo {title} {{Application of
  the theory of dispersion forces to the surface melting of ice}},}\ }\href
  {\doibase 10.1103/PhysRevLett.66.1713} {\bibfield  {journal} {\bibinfo
  {journal} {Phys. Rev. Lett.}\ }\textbf {\bibinfo {volume} {66}},\ \bibinfo
  {pages} {1713--1716} (\bibinfo {year} {1991}{\natexlab{b}})}\BibitemShut
  {NoStop}%
\bibitem [{\citenamefont {Parashar}\ \emph {et~al.}(2019)\citenamefont
  {Parashar}, \citenamefont {Shajesh}, \citenamefont {Milton}, \citenamefont
  {Parsons}, \citenamefont {Brevik},\ and\ \citenamefont
  {Bostr\"om}}]{Prachi_concentricice2019}%
  \BibitemOpen
  \bibfield  {author} {\bibinfo {author} {\bibfnamefont {P.}~\bibnamefont
  {Parashar}}, \bibinfo {author} {\bibfnamefont {K.~V.}\ \bibnamefont
  {Shajesh}}, \bibinfo {author} {\bibfnamefont {K.~A.}\ \bibnamefont {Milton}},
  \bibinfo {author} {\bibfnamefont {D.~F.}\ \bibnamefont {Parsons}}, \bibinfo
  {author} {\bibfnamefont {I.}~\bibnamefont {Brevik}}, \ and\ \bibinfo {author}
  {\bibfnamefont {M.}~\bibnamefont {Bostr\"om}},\ }\bibfield  {title} {\enquote
  {\bibinfo {title} {Role of zero point energy in promoting ice formation in a
  spherical drop of water},}\ }\href {\doibase
  10.1103/PhysRevResearch.1.033210} {\bibfield  {journal} {\bibinfo  {journal}
  {Phys. Rev. Research}\ }\textbf {\bibinfo {volume} {1}},\ \bibinfo {pages}
  {033210} (\bibinfo {year} {2019})}\BibitemShut {NoStop}%
\bibitem [{\citenamefont {Milton}(1980)}]{milton1980semiclassical}%
  \BibitemOpen
  \bibfield  {author} {\bibinfo {author} {\bibfnamefont {Kimball~A}\
  \bibnamefont {Milton}},\ }\bibfield  {title} {\enquote {\bibinfo {title}
  {{Semiclassical electron models: Casimir self-stress in dielectric and
  conducting balls}},}\ }\href@noop {} {\bibfield  {journal} {\bibinfo
  {journal} {Ann. Phys. (N.Y.)}\ }\textbf {\bibinfo {volume} {127}},\ \bibinfo
  {pages} {49--61} (\bibinfo {year} {1980})}\BibitemShut {NoStop}%
\bibitem [{\citenamefont {Candelas}(1982)}]{candelas1982vacuum}%
  \BibitemOpen
  \bibfield  {author} {\bibinfo {author} {\bibfnamefont {Philip}\ \bibnamefont
  {Candelas}},\ }\bibfield  {title} {\enquote {\bibinfo {title} {Vacuum energy
  in the presence of dielectric and conducting surfaces},}\ }\href@noop {}
  {\bibfield  {journal} {\bibinfo  {journal} {Ann. Phys. (N.Y.)}\ }\textbf
  {\bibinfo {volume} {143}},\ \bibinfo {pages} {241--295} (\bibinfo {year}
  {1982})}\BibitemShut {NoStop}%
\bibitem [{\citenamefont {Brevik}\ and\ \citenamefont
  {Kolbenstvedt}(1983)}]{brevik1983electromagnetic}%
  \BibitemOpen
  \bibfield  {author} {\bibinfo {author} {\bibfnamefont {I}~\bibnamefont
  {Brevik}}\ and\ \bibinfo {author} {\bibfnamefont {H}~\bibnamefont
  {Kolbenstvedt}},\ }\bibfield  {title} {\enquote {\bibinfo {title}
  {{Electromagnetic Casimir densities in dielectric spherical media}},}\
  }\href@noop {} {\bibfield  {journal} {\bibinfo  {journal} {Ann. Phys.
  (N.Y.)}\ }\textbf {\bibinfo {volume} {149}},\ \bibinfo {pages} {237--253}
  (\bibinfo {year} {1983})}\BibitemShut {NoStop}%
\bibitem [{\citenamefont {Brevik}\ and\ \citenamefont
  {Nyland}(1994)}]{brevik1994casimir}%
  \BibitemOpen
  \bibfield  {author} {\bibinfo {author} {\bibfnamefont {Iver}\ \bibnamefont
  {Brevik}}\ and\ \bibinfo {author} {\bibfnamefont {G.~H.}\ \bibnamefont
  {Nyland}},\ }\bibfield  {title} {\enquote {\bibinfo {title} {Casimir force on
  a dielectric cylinder},}\ }\href@noop {} {\bibfield  {journal} {\bibinfo
  {journal} {Ann. Phys. (N.Y.)}\ }\textbf {\bibinfo {volume} {230}},\ \bibinfo
  {pages} {321--342} (\bibinfo {year} {1994})}\BibitemShut {NoStop}%
\bibitem [{\citenamefont {Avni}\ and\ \citenamefont
  {Leonhardt}(2018)}]{avni2018casimir}%
  \BibitemOpen
  \bibfield  {author} {\bibinfo {author} {\bibfnamefont {Yael}\ \bibnamefont
  {Avni}}\ and\ \bibinfo {author} {\bibfnamefont {Ulf}\ \bibnamefont
  {Leonhardt}},\ }\bibfield  {title} {\enquote {\bibinfo {title} {Casimir
  self-stress in a dielectric sphere},}\ }\href@noop {} {\bibfield  {journal}
  {\bibinfo  {journal} {Ann. Phys. (N.Y.)}\ }\textbf {\bibinfo {volume}
  {395}},\ \bibinfo {pages} {326--340} (\bibinfo {year} {2018})}\BibitemShut
  {NoStop}%
\bibitem [{\citenamefont {Milton}\ \emph {et~al.}(2020)\citenamefont {Milton},
  \citenamefont {Parashar}, \citenamefont {Brevik},\ and\ \citenamefont
  {Kennedy}}]{milton2020self}%
  \BibitemOpen
  \bibfield  {author} {\bibinfo {author} {\bibfnamefont {Kimball~A}\
  \bibnamefont {Milton}}, \bibinfo {author} {\bibfnamefont {Prachi}\
  \bibnamefont {Parashar}}, \bibinfo {author} {\bibfnamefont {Iver}\
  \bibnamefont {Brevik}}, \ and\ \bibinfo {author} {\bibfnamefont {Gerard}\
  \bibnamefont {Kennedy}},\ }\bibfield  {title} {\enquote {\bibinfo {title}
  {{Self-stress on a dielectric ball and Casimir--Polder forces}},}\
  }\href@noop {} {\bibfield  {journal} {\bibinfo  {journal} {Ann. Phys.
  (N.Y.)}\ }\textbf {\bibinfo {volume} {412}},\ \bibinfo {pages} {168008}
  (\bibinfo {year} {2020})}\BibitemShut {NoStop}%
\bibitem [{\citenamefont {Milton}\ and\ \citenamefont
  {Brevik}(2018)}]{milton2018casimir}%
  \BibitemOpen
  \bibfield  {author} {\bibinfo {author} {\bibfnamefont {Kimball~A}\
  \bibnamefont {Milton}}\ and\ \bibinfo {author} {\bibfnamefont {Iver}\
  \bibnamefont {Brevik}},\ }\bibfield  {title} {\enquote {\bibinfo {title}
  {Casimir energies for isorefractive or diaphanous balls},}\ }\href@noop {}
  {\bibfield  {journal} {\bibinfo  {journal} {Symmetry}\ }\textbf {\bibinfo
  {volume} {10}},\ \bibinfo {pages} {68} (\bibinfo {year} {2018})}\BibitemShut
  {NoStop}%
\bibitem [{\citenamefont {Li}(2024)}]{li2024casimir}%
  \BibitemOpen
  \bibfield  {author} {\bibinfo {author} {\bibfnamefont {Yang}\ \bibnamefont
  {Li}},\ }\bibfield  {title} {\enquote {\bibinfo {title} {Casimir stresses of
  the dielectric ball: inhomogeneity and divergences},}\ }\href@noop {}
  {\bibfield  {journal} {\bibinfo  {journal} {arXiv preprint arXiv:2411.07911}\
  } (\bibinfo {year} {2024})}\BibitemShut {NoStop}%
\bibitem [{\citenamefont {Lifshitz}(1956)}]{lifshitz1956theory}%
  \BibitemOpen
  \bibfield  {author} {\bibinfo {author} {\bibfnamefont {E.~M.}\ \bibnamefont
  {Lifshitz}},\ }\bibfield  {title} {\enquote {\bibinfo {title} {The theory of
  molecular attractive forces between solids},}\ }\href@noop {} {\bibfield
  {journal} {\bibinfo  {journal} {Sov. Phys. JETP}\ }\textbf {\bibinfo {volume}
  {2}},\ \bibinfo {pages} {73} (\bibinfo {year} {1956})}\BibitemShut {NoStop}%
\bibitem [{\citenamefont {Dzyaloshinskii}\ \emph {et~al.}(1961)\citenamefont
  {Dzyaloshinskii}, \citenamefont {Lifshitz},\ and\ \citenamefont
  {Pitaevskii}}]{dzyaloshinskii1961general}%
  \BibitemOpen
  \bibfield  {author} {\bibinfo {author} {\bibfnamefont {I.~E.}\ \bibnamefont
  {Dzyaloshinskii}}, \bibinfo {author} {\bibfnamefont {E.~M.}\ \bibnamefont
  {Lifshitz}}, \ and\ \bibinfo {author} {\bibfnamefont {L.~P.}\ \bibnamefont
  {Pitaevskii}},\ }\bibfield  {title} {\enquote {\bibinfo {title} {{The general
  theory of van der Waals forces}},}\ }\href@noop {} {\bibfield  {journal}
  {\bibinfo  {journal} {Adv. Phys.}\ }\textbf {\bibinfo {volume} {10}},\
  \bibinfo {pages} {165} (\bibinfo {year} {1961})}\BibitemShut {NoStop}%
\bibitem [{\citenamefont {Bostr{\"o}m}\ \emph {et~al.}(2021)\citenamefont
  {Bostr{\"o}m}, \citenamefont {Esteso}, \citenamefont {Fiedler}, \citenamefont
  {Brevik}, \citenamefont {Buhmann}, \citenamefont {Persson}, \citenamefont
  {Carretero-Palacios}, \citenamefont {Parsons},\ and\ \citenamefont
  {Corkery}}]{bostrom2021self}%
  \BibitemOpen
  \bibfield  {author} {\bibinfo {author} {\bibfnamefont {M.}~\bibnamefont
  {Bostr{\"o}m}}, \bibinfo {author} {\bibfnamefont {V.}~\bibnamefont {Esteso}},
  \bibinfo {author} {\bibfnamefont {J.}~\bibnamefont {Fiedler}}, \bibinfo
  {author} {\bibfnamefont {I.}~\bibnamefont {Brevik}}, \bibinfo {author}
  {\bibfnamefont {S.~Y.}\ \bibnamefont {Buhmann}}, \bibinfo {author}
  {\bibfnamefont {C.}~\bibnamefont {Persson}}, \bibinfo {author} {\bibfnamefont
  {S.}~\bibnamefont {Carretero-Palacios}}, \bibinfo {author} {\bibfnamefont
  {D.~F.}\ \bibnamefont {Parsons}}, \ and\ \bibinfo {author} {\bibfnamefont
  {R.~W.}\ \bibnamefont {Corkery}},\ }\bibfield  {title} {\enquote {\bibinfo
  {title} {{Self-preserving ice layers on $\rm CO_2$ clathrate particles:
  Implications for Enceladus, Pluto, and similar ocean worlds}},}\ }\href@noop
  {} {\bibfield  {journal} {\bibinfo  {journal} {Astron. Astrophys.}\ }\textbf
  {\bibinfo {volume} {650}},\ \bibinfo {pages} {A54} (\bibinfo {year}
  {2021})}\BibitemShut {NoStop}%
\bibitem [{\citenamefont {Kerker}(1969)}]{Kerker}%
  \BibitemOpen
  \bibfield  {author} {\bibinfo {author} {\bibfnamefont {M.}~\bibnamefont
  {Kerker}},\ }\href {https://doi.org/10.1016/C2013-0-06195-6} {\emph {\bibinfo
  {title} {{The scattering of light and electromagnetic radiation}}}}\
  (\bibinfo  {publisher} {Academic Press},\ \bibinfo {address} {New York},\
  \bibinfo {year} {1969})\BibitemShut {NoStop}%
\bibitem [{\citenamefont {Esteso}\ \emph {et~al.}(2020)\citenamefont {Esteso},
  \citenamefont {Carretero-Palacios}, \citenamefont {MacDowell}, \citenamefont
  {Fiedler}, \citenamefont {Parsons}, \citenamefont {Spallek}, \citenamefont
  {M\'iguez}, \citenamefont {Persson}, \citenamefont {Buhmann}, \citenamefont
  {Brevik},\ and\ \citenamefont {Bostr\"om}}]{Esteso4layerPCCP2020}%
  \BibitemOpen
  \bibfield  {author} {\bibinfo {author} {\bibfnamefont {V.}~\bibnamefont
  {Esteso}}, \bibinfo {author} {\bibfnamefont {S.}~\bibnamefont
  {Carretero-Palacios}}, \bibinfo {author} {\bibfnamefont {L.~G.}\ \bibnamefont
  {MacDowell}}, \bibinfo {author} {\bibfnamefont {J.}~\bibnamefont {Fiedler}},
  \bibinfo {author} {\bibfnamefont {D.~F.}\ \bibnamefont {Parsons}}, \bibinfo
  {author} {\bibfnamefont {F.}~\bibnamefont {Spallek}}, \bibinfo {author}
  {\bibfnamefont {H.}~\bibnamefont {M\'iguez}}, \bibinfo {author}
  {\bibfnamefont {C.}~\bibnamefont {Persson}}, \bibinfo {author} {\bibfnamefont
  {S.~Y.}\ \bibnamefont {Buhmann}}, \bibinfo {author} {\bibfnamefont
  {I.}~\bibnamefont {Brevik}}, \ and\ \bibinfo {author} {\bibfnamefont
  {M.}~\bibnamefont {Bostr\"om}},\ }\bibfield  {title} {\enquote {\bibinfo
  {title} {Premelting of ice adsorbed on a rock surface},}\ }\href {\doibase
  10.1039/C9CP06836H} {\bibfield  {journal} {\bibinfo  {journal} {Phys. Chem.
  Chem. Phys.}\ }\textbf {\bibinfo {volume} {22}},\ \bibinfo {pages}
  {11362--11373} (\bibinfo {year} {2020})}\BibitemShut {NoStop}%
\bibitem [{\citenamefont {Li}\ \emph {et~al.}(2022)\citenamefont {Li},
  \citenamefont {Milton}, \citenamefont {Brevik}, \citenamefont {Malyi},
  \citenamefont {Thiyam}, \citenamefont {Persson}, \citenamefont {Parsons},\
  and\ \citenamefont
  {Bostr\"om}}]{LiMiltonBrevikMalyiThiyamPerssonParsonsBostrom_PRB2022}%
  \BibitemOpen
  \bibfield  {author} {\bibinfo {author} {\bibfnamefont {Y.}~\bibnamefont
  {Li}}, \bibinfo {author} {\bibfnamefont {K.~A.}\ \bibnamefont {Milton}},
  \bibinfo {author} {\bibfnamefont {I.}~\bibnamefont {Brevik}}, \bibinfo
  {author} {\bibfnamefont {O.~I.}\ \bibnamefont {Malyi}}, \bibinfo {author}
  {\bibfnamefont {P.}~\bibnamefont {Thiyam}}, \bibinfo {author} {\bibfnamefont
  {C.}~\bibnamefont {Persson}}, \bibinfo {author} {\bibfnamefont {D.~F.}\
  \bibnamefont {Parsons}}, \ and\ \bibinfo {author} {\bibfnamefont
  {M.}~\bibnamefont {Bostr\"om}},\ }\bibfield  {title} {\enquote {\bibinfo
  {title} {Premelting and formation of ice due to {Casimir-Lifshitz}
  interactions: Impact of improved parameterization for materials},}\ }\href
  {\doibase 10.1103/PhysRevB.105.014203} {\bibfield  {journal} {\bibinfo
  {journal} {Phys. Rev. B}\ }\textbf {\bibinfo {volume} {105}},\ \bibinfo
  {pages} {014203} (\bibinfo {year} {2022})}\BibitemShut {NoStop}%
\bibitem [{\citenamefont {Thiyam}\ \emph {et~al.}(2018)\citenamefont {Thiyam},
  \citenamefont {Fiedler}, \citenamefont {Buhmann}, \citenamefont {Persson},
  \citenamefont {Brevik}, \citenamefont {Bostr{\"o}m},\ and\ \citenamefont
  {Parsons}}]{ThiyamFiedlerBuhmannPerssonBrevikBostromParsons2018}%
  \BibitemOpen
  \bibfield  {author} {\bibinfo {author} {\bibfnamefont {P.}~\bibnamefont
  {Thiyam}}, \bibinfo {author} {\bibfnamefont {J.}~\bibnamefont {Fiedler}},
  \bibinfo {author} {\bibfnamefont {S.~Y.}\ \bibnamefont {Buhmann}}, \bibinfo
  {author} {\bibfnamefont {C.}~\bibnamefont {Persson}}, \bibinfo {author}
  {\bibfnamefont {I.}~\bibnamefont {Brevik}}, \bibinfo {author} {\bibfnamefont
  {M.}~\bibnamefont {Bostr{\"o}m}}, \ and\ \bibinfo {author} {\bibfnamefont
  {D.~F.}\ \bibnamefont {Parsons}},\ }\bibfield  {title} {\enquote {\bibinfo
  {title} {{Ice Particles Sink below the Water Surface Due to a Balance of
  Salt, van der Waals, and Buoyancy Forces}},}\ }\href {\doibase
  10.1021/acs.jpcc.8b02351} {\bibfield  {journal} {\bibinfo  {journal} {J.
  Phys. Chem. C}\ }\textbf {\bibinfo {volume} {122}},\ \bibinfo {pages}
  {15311--15317} (\bibinfo {year} {2018})}\BibitemShut {NoStop}%
\bibitem [{\citenamefont {Stoiber}\ \emph {et~al.}(1987)\citenamefont
  {Stoiber}, \citenamefont {Williams},\ and\ \citenamefont
  {Huebert}}]{STOIBER1987}%
  \BibitemOpen
  \bibfield  {author} {\bibinfo {author} {\bibfnamefont {R.~E.}\ \bibnamefont
  {Stoiber}}, \bibinfo {author} {\bibfnamefont {S.~N.}\ \bibnamefont
  {Williams}}, \ and\ \bibinfo {author} {\bibfnamefont {B.}~\bibnamefont
  {Huebert}},\ }\bibfield  {title} {\enquote {\bibinfo {title} {Annual
  contribution of sulfur dioxide to the atmosphere by volcanoes},}\ }\href
  {\doibase https://doi.org/10.1016/0377-0273(87)90051-5} {\bibfield  {journal}
  {\bibinfo  {journal} {J. Volcanol. Geotherm. Res.}\ }\textbf {\bibinfo
  {volume} {33}},\ \bibinfo {pages} {1--8} (\bibinfo {year}
  {1987})}\BibitemShut {NoStop}%
\bibitem [{\citenamefont {Parsons}\ and\ \citenamefont
  {Salis}(2015)}]{ParsonsSalis2015}%
  \BibitemOpen
  \bibfield  {author} {\bibinfo {author} {\bibfnamefont {D.~F.}\ \bibnamefont
  {Parsons}}\ and\ \bibinfo {author} {\bibfnamefont {A.}~\bibnamefont
  {Salis}},\ }\bibfield  {title} {\enquote {\bibinfo {title} {The impact of the
  competitive adsorption of ions at surface sites on surface free energies and
  surface forces},}\ }\href {\doibase 10.1063/1.4916519} {\bibfield  {journal}
  {\bibinfo  {journal} {J. Chem. Phys.}\ }\textbf {\bibinfo {volume} {142}},\
  \bibinfo {pages} {134707} (\bibinfo {year} {2015})}\BibitemShut {NoStop}%
\bibitem [{\citenamefont {Parsons}\ \emph {et~al.}(2022)\citenamefont
  {Parsons}, \citenamefont {Carucci},\ and\ \citenamefont
  {Salis}}]{Parsons2022}%
  \BibitemOpen
  \bibfield  {author} {\bibinfo {author} {\bibfnamefont {D.~F.}\ \bibnamefont
  {Parsons}}, \bibinfo {author} {\bibfnamefont {C.}~\bibnamefont {Carucci}}, \
  and\ \bibinfo {author} {\bibfnamefont {A.}~\bibnamefont {Salis}},\ }\bibfield
   {title} {\enquote {\bibinfo {title} {{Buffer-specific effects arise from
  ionic dispersion forces}},}\ }\href {\doibase 10.1039/D2CP00223J} {\bibfield
  {journal} {\bibinfo  {journal} {Phys. Chem. Chem. Phys.}\ }\textbf {\bibinfo
  {volume} {24}},\ \bibinfo {pages} {6544--6551} (\bibinfo {year}
  {2022})}\BibitemShut {NoStop}%
\bibitem [{\citenamefont {Pruppacher}\ and\ \citenamefont
  {Klett}(2010)}]{PruppacherKlett}%
  \BibitemOpen
  \bibfield  {author} {\bibinfo {author} {\bibfnamefont {H.~R.}\ \bibnamefont
  {Pruppacher}}\ and\ \bibinfo {author} {\bibfnamefont {J.~D.}\ \bibnamefont
  {Klett}},\ }\href {\doibase 10.1007/978-0-306-48100-0} {\emph {\bibinfo
  {title} {Microphysics of Clouds and Precipitation}}},\ {Atmospheric and
  Oceanographic Sciences Library}\ (\bibinfo  {publisher} {Springer
  Dordrecht},\ \bibinfo {year} {2010})\BibitemShut {NoStop}%
\bibitem [{\citenamefont {Knopf}\ and\ \citenamefont
  {Alpert}(2023)}]{KnopfAlpertNatureRevPhys2023}%
  \BibitemOpen
  \bibfield  {author} {\bibinfo {author} {\bibfnamefont {D.~A.}\ \bibnamefont
  {Knopf}}\ and\ \bibinfo {author} {\bibfnamefont {P.~A.}\ \bibnamefont
  {Alpert}},\ }\bibfield  {title} {\enquote {\bibinfo {title} {{Atmospheric ice
  nucleation}},}\ }\href {\doibase 10.1038/s42254-023-00570-7} {\bibfield
  {journal} {\bibinfo  {journal} {Nat. Rev. Phys.}\ }\textbf {\bibinfo {volume}
  {5}},\ \bibinfo {pages} {203--217} (\bibinfo {year} {2023})}\BibitemShut
  {NoStop}%
\bibitem [{\citenamefont {Olver}(2010)}]{olver2010nist}%
  \BibitemOpen
  \bibfield  {author} {\bibinfo {author} {\bibfnamefont {F.~W.~J.}\
  \bibnamefont {Olver}},\ }\href@noop {} {\emph {\bibinfo {title} {{NIST
  handbook of mathematical functions hardback and CD-ROM}}}}\ (\bibinfo
  {publisher} {Cambridge university press},\ \bibinfo {year}
  {2010})\BibitemShut {NoStop}%
\end{thebibliography}%

\end{document}